\documentclass[aps,pra,reprint,showpacs,nofootinbib,superscriptaddress]{revtex4-2}
\usepackage{graphicx,comment}
\usepackage{amsfonts}
\usepackage{amsmath,amssymb}
\usepackage{braket}
\usepackage{bm}
\usepackage{color}
\usepackage{epstopdf}
\usepackage{epsfig}
\usepackage[font=small,
   justification=justified,
   format=plain]{caption} 
\usepackage{float}
\usepackage{verbatim}
\usepackage{hyperref}
\usepackage{multirow}
\usepackage[table]{xcolor}
\usepackage{wrapfig}
\usepackage{ulem}
\usepackage{natbib}
\usepackage{caption,subcaption}

\hypersetup{
     colorlinks   = true,
     citecolor    = blue
}
\usepackage{lineno}
\usepackage[thicklines]{cancel}
\usepackage{color, colortbl}
\definecolor{LightCyan}{rgb}{0.88,1,1}
\usepackage{url}
\usepackage{dcolumn}
    
{\rm }

\def\be{\begin{equation}}
 \def\ee{\end{equation}}
 \def\bea{\begin{eqnarray}}
 \def\eea{\end{eqnarray}}

\def\2{\frac{1}{2}}

\def\4{\frac{1}{4}}

\begin{document}
\title{Experimentally Realizable Continuous-variable Quantum Neural Networks}

\author{Shikha Bangar}
\email{sbangar@vols.utk.edu}
\affiliation{Department of Physics and Astronomy,  The University of Tennessee, Knoxville, TN 37996-1200, USA}

\author{Leanto Sunny}
\email{lsunny@vols.utk.edu}
\affiliation{Department of Physics and Astronomy,  The University of Tennessee, Knoxville, TN 37996-1200, USA}

\author{K\"ubra Yeter-Aydeniz}
\email{kyeteraydeniz@mitre.org}
\affiliation{Emerging Technologies and Physical Sciences Department, The MITRE Corporation, 7515 Colshire Drive, McLean, Virginia 22102-7539, USA}

\author{George Siopsis}
\email{siopsis@tennessee.edu}
\affiliation{Department of Physics and Astronomy,  The University of Tennessee, Knoxville, TN 37996-1200, USA}

\date{\today}

\begin{abstract}
Continuous-variable (CV) quantum computing has shown great potential for building neural network models. These neural networks can have different levels of quantum-classical hybridization depending on the complexity of the problem. Previous work on CV neural network protocols required the implementation of non-Gaussian operators in the network. These operators were used to introduce non-linearity, an essential feature of neural networks. However, these protocols are hard to execute experimentally. We built a CV hybrid quantum-classical neural network protocol that can be realized experimentally with current photonic quantum hardware. Our protocol uses Gaussian gates only with the addition of ancillary qumodes. We implemented non-linearity through repeat-until-success measurements on ancillary qumodes. To test our neural network, we studied canonical machine learning and quantum computer problems in a supervised learning setting -- state preparation, curve fitting, and classification problems. We achieved high fidelity in state preparation of single-photon (99.9\%), cat (99.8\%), and Gottesman-Kitaev-Preskill  (93.9\%) states, a well-fitted curve in the presence of noise at a cost of less than 1\%, and more than 95\% accuracy in classification problems. These results bode well for real-world applications of CV quantum neural networks.

\end{abstract}

\maketitle

\section{Introduction}
Continuous-variable (CV) quantum computing (QC) takes advantage of the wave-like properties of particles. For example, it can be realized by photonic quantum hardware manipulating electromagnetic fields. Thus, CV QC is achievable in quantum optics by utilizing continuous quadratures of the quantized electromagnetic field \cite{RevModPhys.77.513} enabling the essential steps in quantum algorithms (preparation, unitary manipulation, and measurement of (entangled) quantum states). CV quantum algorithms have been developed for various applications, ranging from quantum field theory \cite{PhysRevA.92.063825,PhysRevD.97.036004,PhysRevA.105.012412} to machine learning \cite{PhysRevLett.118.080501}.
It has recently been shown that CV is also a good architecture for building quantum neural network (QNN) models on quantum computers \cite{PhysRevResearch.1.033063}. The developed CV QNN architecture has been applied to various practical real-world problems, e.g., function fitting, fraud detection of credit card transactions, image classification of hand-written digits, data encryption and decryption in a secure cryptography algorithm \cite{Shi2020}, and entangled state detection \cite{8764462}. In addition to these applications, CV QNN has been utilized as generator and discriminator in a CV quantum adversarial network (CV QGAN) \cite{chang2021quantum} to reproduce the data outputs of the calorimeters for data collected in high energy physics experiments at CERN. 

In the CV QNN model discussed in \cite{PhysRevResearch.1.033063}, a non-Gaussian Kerr gate provides the non-linearity required for neural networks. However, experimentally realizing these non-Gaussian operations is challenging due to their weakly interacting nature. Therefore, we developed an alternative CV QNN model in which non-linearities are introduced through measurements on ancillary qumodes following the proposal in \cite{marshall2015repeat} based on repeat-until-success measurements. In this paper, we propose a variational hybrid quantum-classical circuit implementing a neural network that solves well-known problems such as function fitting, state preparation, binary classification, and image recognition.   
The quantum circuit uses only Gaussian gates that can be implemented with optical elements, such as beam splitters and squeezers. Thus, our CV hybrid neural network can be realized experimentally using current photonic quantum hardware.   

Similar measurements on ancillary qumodes as a means towards creating desired quantum states have been considered before. Single- and two-mode quantum gates acting on photonic qubits were generated with single photon sources, linear optical elements and measurements of ancillary modes \cite{PhysRevA.68.032310}. To generate the cubic phase gate, nonlinear quadrature measurements were implemented using ancillary states, homodyne measurements, and nonlinear feedforwards based on the measurement results \cite{PhysRevApplied.15.024024}. Photon-number-resolving (PNR) measurements were used for the probabilistic production of multi-mode Gaussian states, including cat (superpositions of coherent states), ON, Gottesman-Kitaev-Preskill (GKP) \cite{PhysRevA.64.012310}, NOON \cite{PhysRevA.65.052104}, and bosonic code states \cite{PhysRevA.100.052301}. Cat and GKP states were also created experimentally within a CV cluster state by performing PNR measurements  \cite{Eaton2022measurementbased}. To the best of our knowledge, ours is the first attempt at applying non-linearities induced by measurements to CV QNNs.

The structure of this paper is as follows. We begin with a description of the CV neural network in Section \ref{section:methods}, followed by a detailed account of our proposal to introduce non-linearities that avoids non-Gaussian gates. In Section \ref{section:exp}, we study several problems involving varying degrees of hybridization between quantum and classical neural networks. We conclude with a discussion of the potential of this work in Section \ref{section:conclude}.

\section{The Method} \label{section:methods}
\begin{figure}[ht!]
    \centering
    \begin{subfigure}{.4\textwidth}
\    \includegraphics[scale=0.7,  trim={3cm 0cm 3cm 0cm}]{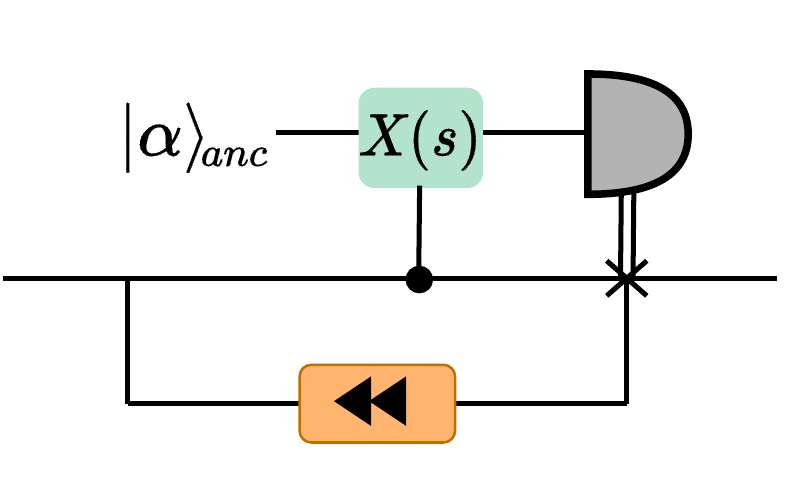}
    \caption{Nonlinear circuit element}
    \end{subfigure}
    \begin{subfigure}{.4\textwidth}
    \includegraphics[scale=0.5,  trim={3cm 0cm 3cm 0cm}]{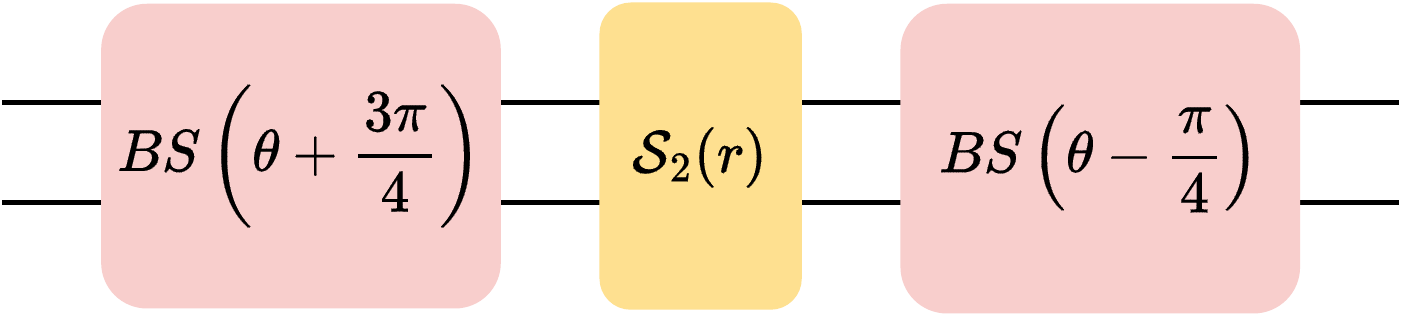}
    \caption{The $CX(s)$ gate }
\end{subfigure}
    \caption{Two-qumode Gaussian quantum circuit element introducing non-linearity in a CV QNN. (a) The ancillary qumode, initially in the coherent state $|\alpha\rangle_{\mathrm{anc}} = \mathcal{D}(\alpha) \ket{0}_{\mathrm{anc}}$, is entangled with the primary qumode with a controlled displacement ($CX (s)$) gate. The primary qumode goes through a feedback loop until the detector on the ancillary qumode clicks \cite{marshall2015repeat}. (b) The $CX(s)$ gate with parameters given in Eq.\ \eqref{eq:CX}. }
    \label{fig:nonlin}
\end{figure}

A general CV QNN was discussed in Ref.\ \cite{PhysRevResearch.1.033063}. It included $N$-port linear optical interferometers consisting of beam splitter ($\mathcal{BS}$), rotation ($\mathcal{R}$), displacement  ($\mathcal{D}$), and squeezing ($\mathcal{S}$) gates. It also featured a non-Gaussian Kerr gate ($\Phi$) introducing non-linearity to the neural network. In our setup, we have replaced this single-mode gate with a two-mode quantum circuit element consisting of Gaussian gates and a photon detector which is experimentally feasible with current technology (shown in Fig.~\ref{fig:nonlin}).

In more detail, our nonlinear circuit element is implemented by adding an ancilla qumode in a coherent state $\ket{\alpha} = \mathcal{D} (\alpha) \ket{0}_{\text{anc}}$ to the primary qumode, where $\alpha\in \mathbb{R}$. The two modes are then entangled by a controlled displacement ($CX(s)$) gate that uses the $q$-quadrature of the primary mode to shift the $q_{\text{anc}}$ quadrature by $sq$, where $s\in \mathbb{R}$. It is implemented with beam splitters and a two-mode squeezer of parameters
\be\label{eq:CX} \cot 2\theta = \sinh r = - s \ . \ee

If the incoming state of the primary mode is $\ket{\psi} = \int dq\, \psi(q) \ket{q}$, then the final two-mode state is the entangled state \be \int dq\, \psi(q) \ket{\alpha + sq}_{\text{anc}} \otimes\ket{q}\ . \ee
Then we use a photon detector to measure the photon number of the ancilla qumode. If the detector clicks, the process is considered successful, and the primary qumode proceeds to the next layer. For small $\alpha$ and $s$ parameters, the ancilla qumode decouples and the outgoing primary qumode is in the (unnormalized) state
\be \int dq\, (\alpha + sq) \psi(q)  \ket{q}\ , \ee
showing that we have effectively applied the non-Gaussian gate $\alpha + sq$. For larger values of the parameters, the expression for this effective gate is more complicated, but it is still a non-Gaussian gate.

If there is no click in the detector, the output state of the primary mode is approximately the same as the input state, $\ket{\psi}$ for small $\alpha$ and $s$, changing slightly for larger values of the parameters (with insertions of factors of $e^{-|\alpha +sq|^2/2}$). We feed the state back into the input port and repeat the process shown in Fig.\ \ref{fig:nonlin}. This loop continues until the detector clicks and the primary qumode can advance (repeat-until-success process \cite{marshall2015repeat}). We fix the coherence parameter $\alpha$ of the ancillary qumode, treating it as a hyperparameter and not a trainable parameter. In our simulations, we set $\alpha=1$. 

\begin{figure}[ht!]
    \centering
    \begin{subfigure}{.4\textwidth}
     \includegraphics[scale=0.4]{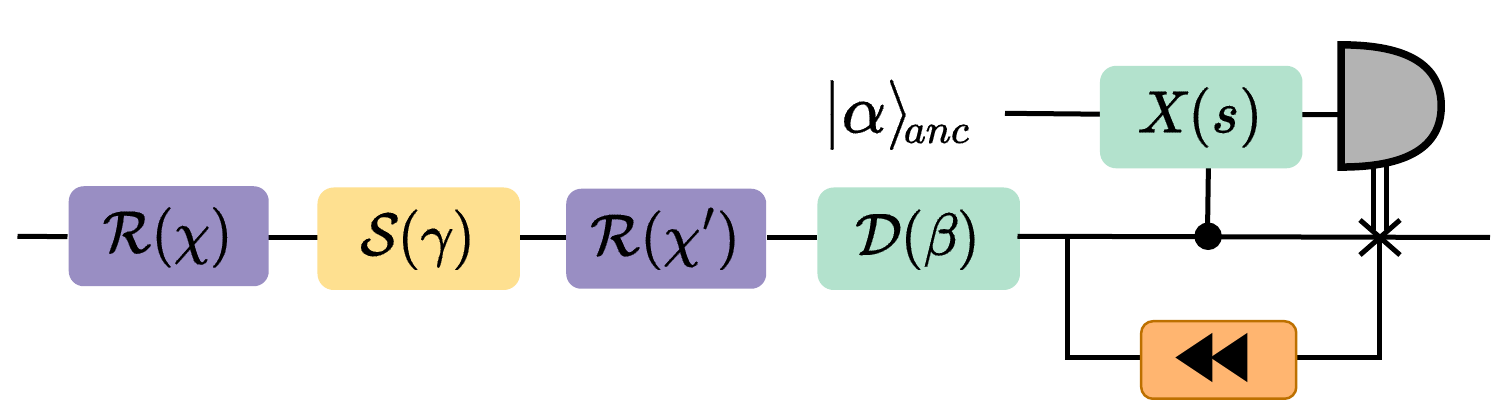}
     \caption{Single-mode layer}
     \label{fig:modenna}
     \end{subfigure}
     
    \begin{subfigure}{.4\textwidth}
     \includegraphics[scale=0.35]{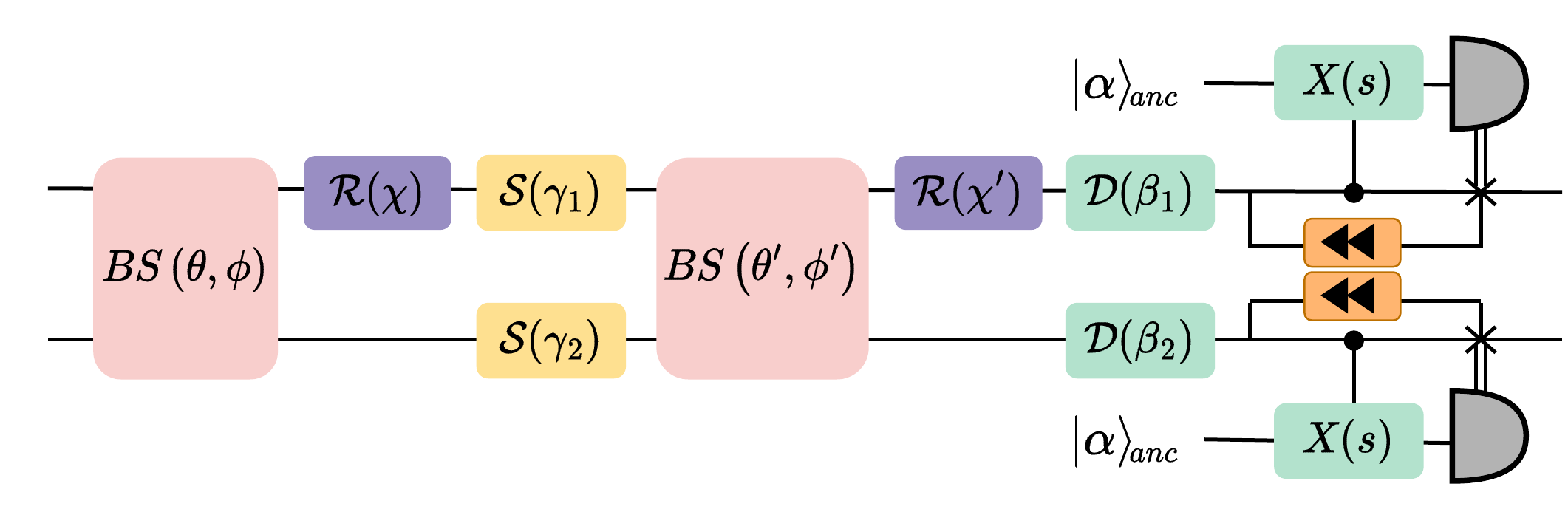}
\caption{Two-mode layer}
\label{fig:modennb}
     \end{subfigure}
    \caption{Detailed CV neural network architecture for (a) single-mode and (b) two-mode layer. }
    \label{fig:modenn}
\end{figure}

A detailed neural network architecture for single-mode and two-mode layers is shown in Fig.~\ref{fig:modenn}. We create a multiple-layer structure by arranging each layer as a building block of the neural network with the gate variables ($\theta, \phi, \chi, x, \alpha, \beta$) being free parameters, collectively denoted by $\vec{\zeta}$. We want to find $\vec{\zeta}$ such that the value of the cost function ($C(\vec{\zeta})$) is minimum. This can be done using various optimization techniques available in deep learning (for example, gradient descent, stochastic gradient descent, and, most commonly, the Adam optimizer). The parameters are updated by the rule: 
\begin{equation}
    \vec{\zeta} \longrightarrow \vec{\zeta} + \eta \nabla_{\vec{\zeta}} \, C(\vec{\zeta})~,
\end{equation}
where $\eta$ is the step size, also known as the learning rate. This procedure continues until the model cannot modify the cost function further. Then, the parameters associated with the lowest observed cost function offer the best solution to the given task. Note that the free parameters correspond to a circuit; hence, the circuit is the solution.

In principle, only one layer ($\mathcal{L}$) is sufficient to parameterize every possible unitary affine transformation on $N$ modes. 
However, deeper architectures provide increased expressive power, better learning capabilities, and a more efficient representation of complex transformations. 


To test our neural network, we studied various machine learning and quantum computation problems, namely, state preparation, curve fitting, binary classification, and image recognition with varying degrees of hybridization in quantum and classical neural networks, as discussed in  Section \ref{section:exp}. 

We conducted our simulations using Strawberry Fields quantum computing software. The realization of the quantum circuit element implementing non-linearity depicted in Figure \ref{fig:nonlin} in a QNN setting turned out to be challenging. To obtain a good approximation to our setup that could be implemented with software available to us, we adjusted the coherent parameter of the ancillary qumode so that a single photon would be detected in that mode with a high success probability. 
To estimate the number of ancillary qumode measurements (or feedback loops) necessary for the successful application of the quantum circuit element, we utilized Bosonic
Qiskit software \cite{stavenger2022c2qa} for simulating an elementary  circuit and tracking the number of measurements required for success in various applications.

\begin{figure}[ht!]
\centering
    \begin{subfigure}{0.4 \textwidth}
\includegraphics[width=0.9 \textwidth]{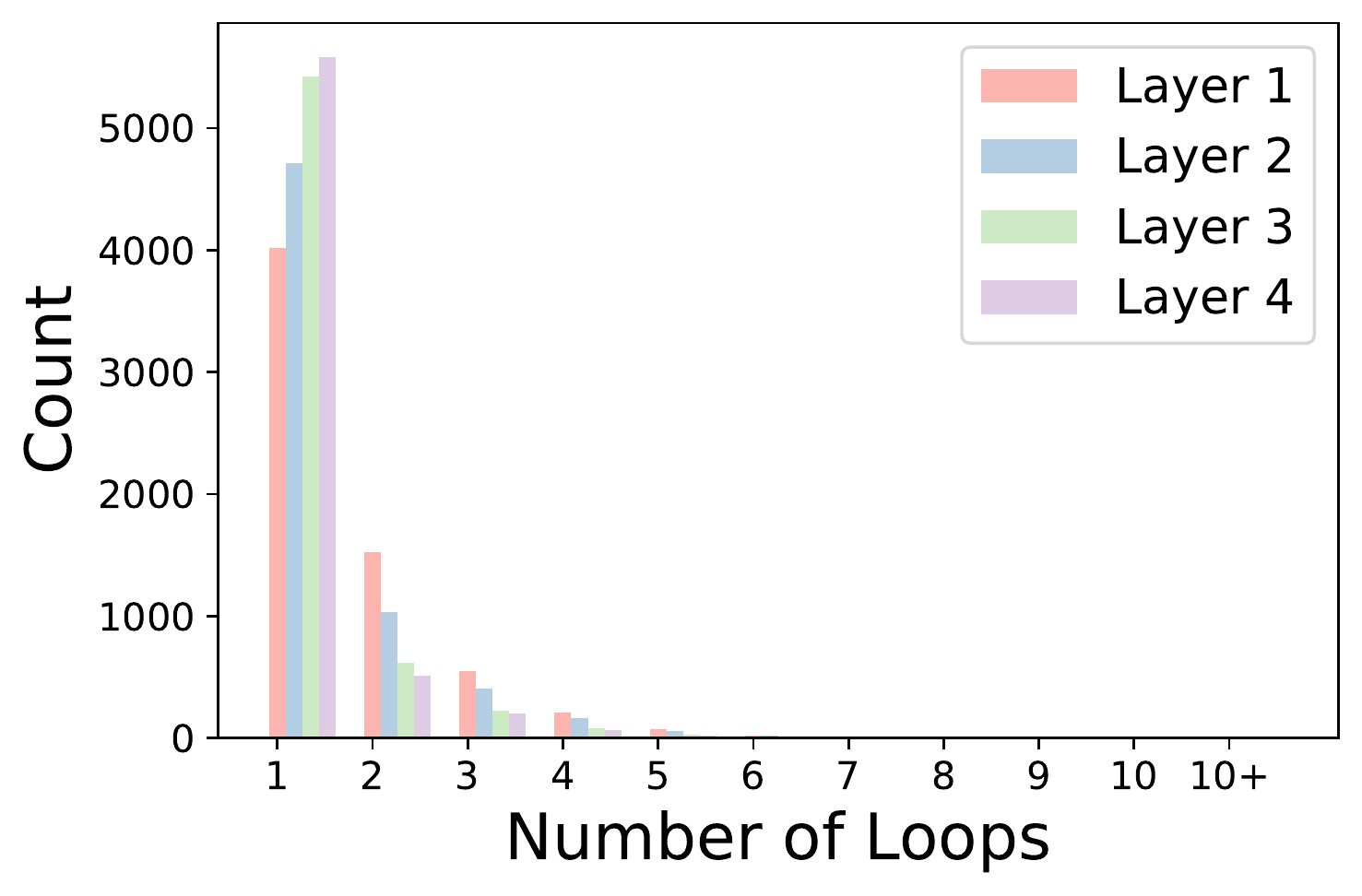}
\caption{4 quantum layers}
\end{subfigure}
    \begin{subfigure}{0.4 \textwidth}
\includegraphics[width=0.9 \textwidth]{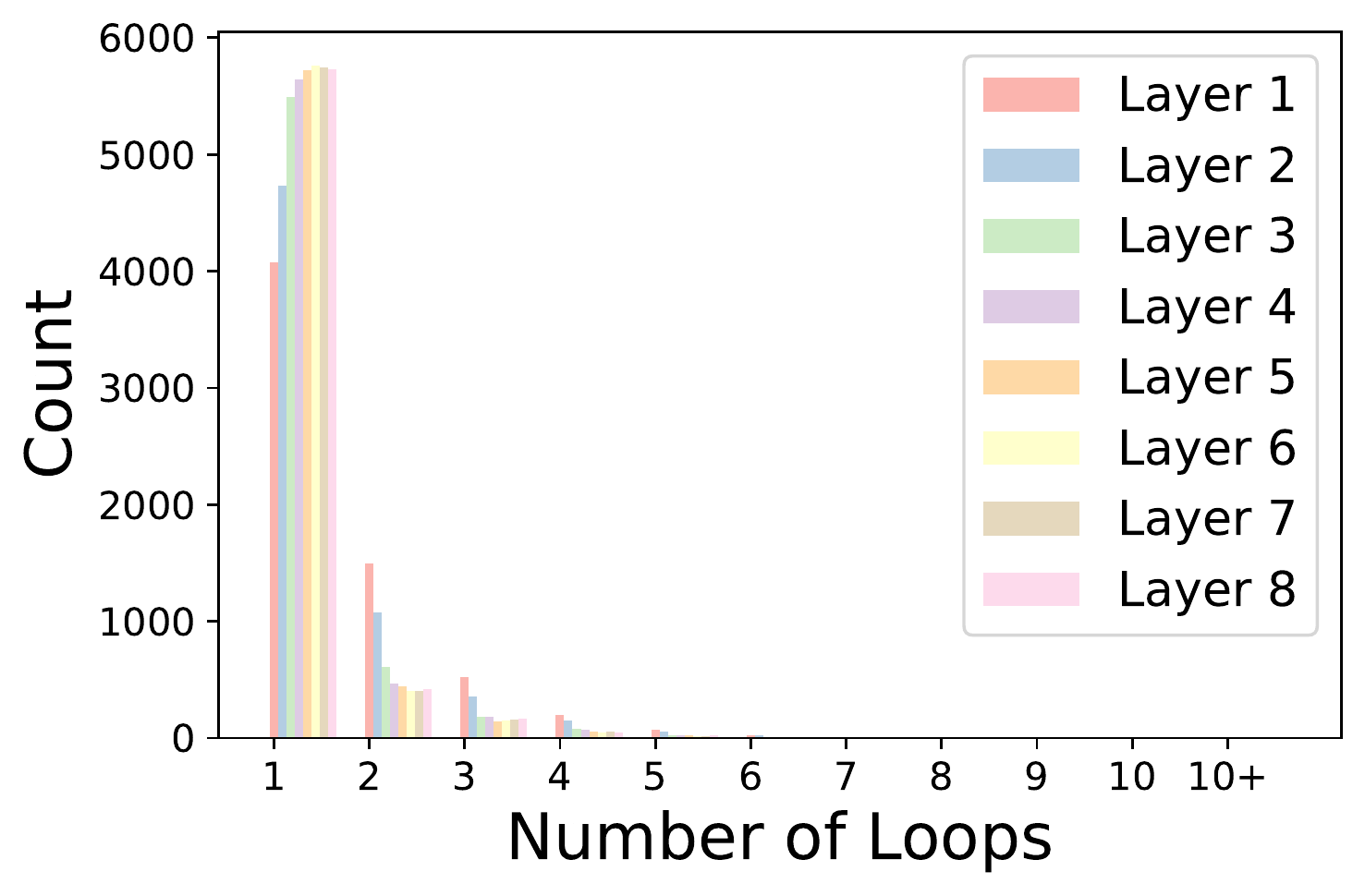}
\caption{8 quantum layers}
\end{subfigure}
\caption{Number of measurements necessary on the ancillary qumode to achieve successful measurement for each quantum layer of sample hybrid networks for binary classification, consisting of two classical and (a) four or (b) eight quantum layers. }\label{fig:loop_count_1}
\end{figure}

Sample results are illustrated in Figure \ref{fig:loop_count_1}. Using the architecture of the binary classification circuit discussed in Section \ref{sec:IIIC} as a concrete example, we counted the number of repeated ancillary qumode measurements required for a successful pass of the primary qumodes during forward propagation through the network. We plotted the success rates per layer for setups utilizing various numbers of layers. Evidently, most successful measurements occur at the first photon detection of the ancillary qumode. 

\begin{figure}
    \begin{subfigure}{0.4 \textwidth}
\centering
\includegraphics[width=0.9 \textwidth]{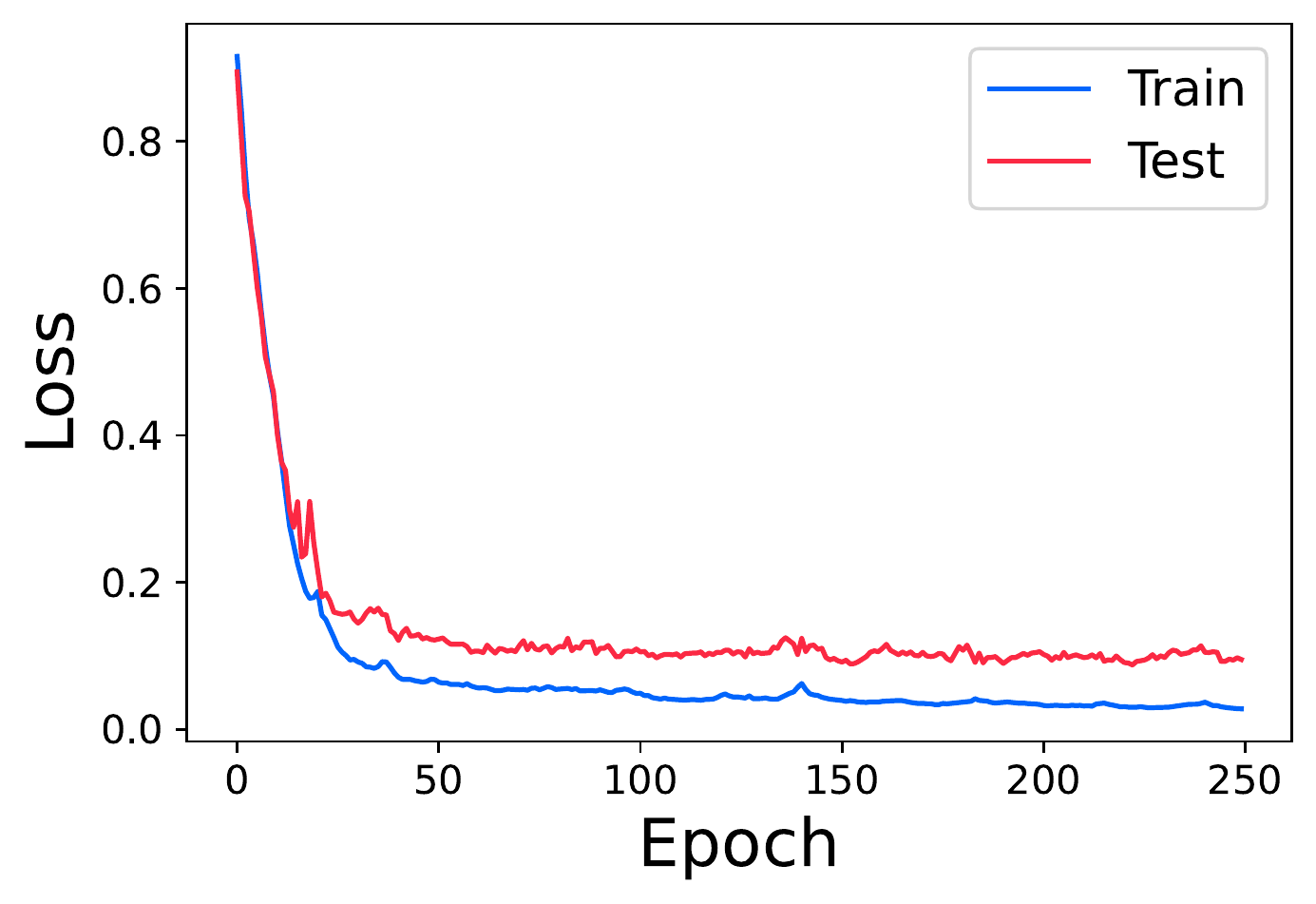}
\caption{1 Feedback Loop}
\end{subfigure}
\centering
    \begin{subfigure}{0.4 \textwidth}
\includegraphics[width=0.9 \textwidth]{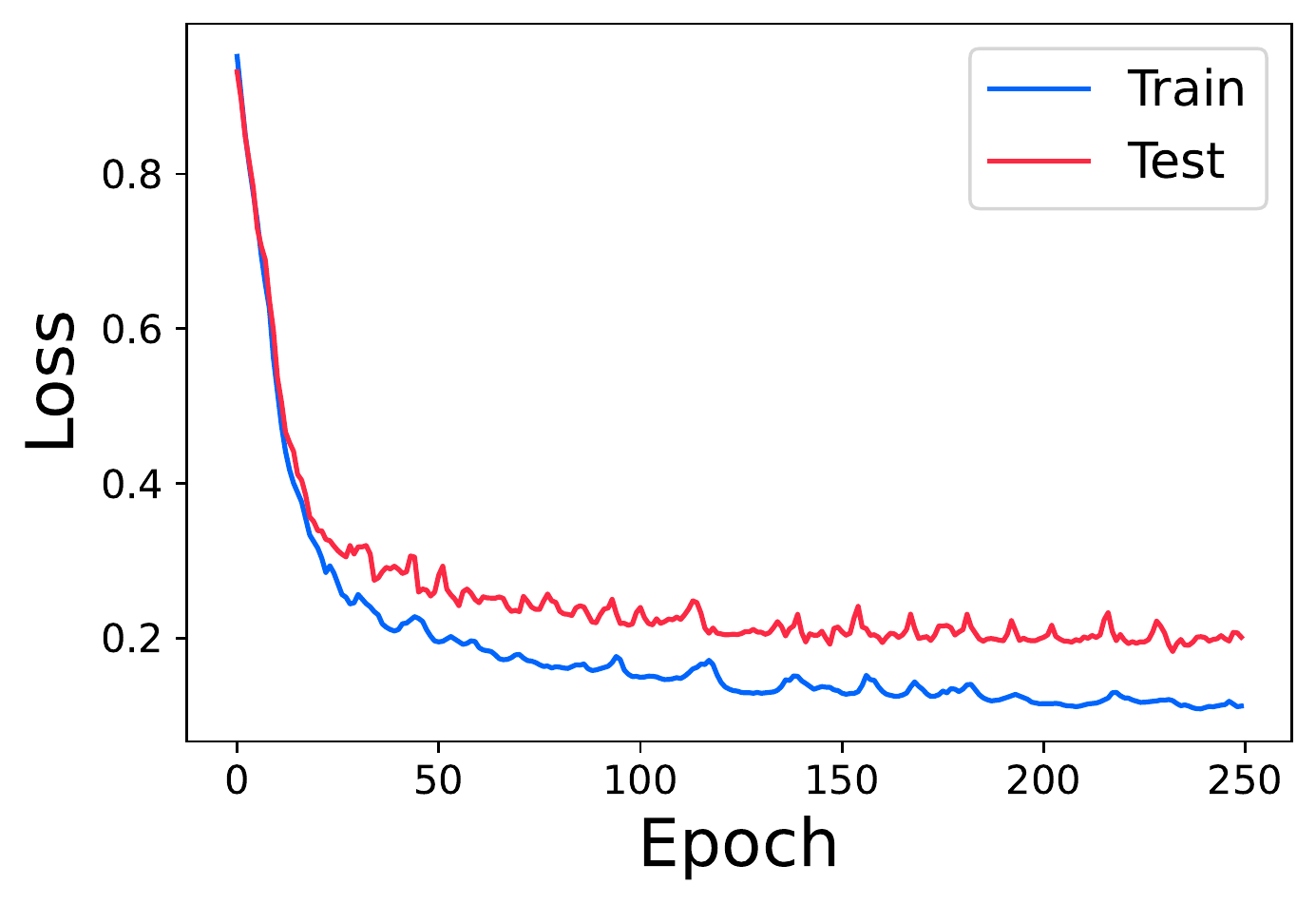}
\caption{2 Feedback Loops}
\end{subfigure}
\caption{Training and testing loss of MNIST classification models with successful measurement of the ancilla within (a) one loop (Model 1) and  (b) two loops (Model 2). }\label{fig:2_loop_loss}
\end{figure}

Due to the constraints in the Strawberry Fields library, we were compelled to perform post-selection of a single Fock state during the ancilla measurement step to collapse the wavefunction, thus effectively implementing meassurements by photon-number-resolving detectors. It would also be interesting to simulate photon detectors that cannot resolve photon number and are widely available. This would alter the effective nonlinear operation slightly (depending on the choice of the laser intensity for the ancillary qumodes) but would simplify the experimental setup. To demonstrate that a high success rate can be achieved even without a photon-number-resolving detector, we trained two distinct models that performed multi-label classification on the MNIST handwritten digit data set discussed in Section \ref{sec:IIID}. The training and testing loss values for these classical-quantum hybrid models are presented in Fig.~\ref{fig:2_loop_loss}. Model 1 was designed to successfully select the required state in the ancilla measurement on the initial attempt whereas Model 2 was set up to fail the first measurement attempt and succeed on the second measurement. Even though the loss function and classification accuracy experienced a slight decline at the end of training from 97.25\% to 96.49\%, we were still able to train Model 2 successfully and achieve a reasonably high classification accuracy for the 4-class MNIST handwritten digit dataset classification.

\section{Case Studies} \label{section:exp}
In this section, we develop CV QNN models for various applications, including quantum state preparation (Section \ref{sec:IIIA}), curve fitting (Section \ref{sec:IIIB}), binary classification of fraud and genuine credit card transactions (Section \ref{sec:IIIC}), and multi-label classification of MNIST handwritten digits (Section \ref{sec:IIID}).
In order to observe the effect of the classical and quantum neural network layers, we analyze both hybrid quantum-classical and fully quantum layers.

We simulated our CV QNN models using the Strawberry Fields software platform \cite{strawberryfields}. The quantum machine learning toolbox application is built on top of it with Tensorflow features \cite{abadi2016tensorflow}. We used the quantum circuit simulator, optimized the algorithm, and trained the neural network to obtain the desired results.

\subsection{State Preparation} \label{sec:IIIA}
The CV QNN model for quantum state preparation trains a quantum circuit to generate a target quantum state. To this end, we provide a canonical input state $\ket{\Psi_0}$ and target output state $\ket{\Psi_t}$ and aim to find out the circuit $U$ (a unitary transformation) such that 
\begin{equation}
    \ket{\Psi_t} =  U \ket{\Psi_0}~.
\end{equation}
\begin{figure}[ht]
\centering
    \includegraphics[scale=0.44,  trim={3cm 3.5cm 4cm 0cm}]{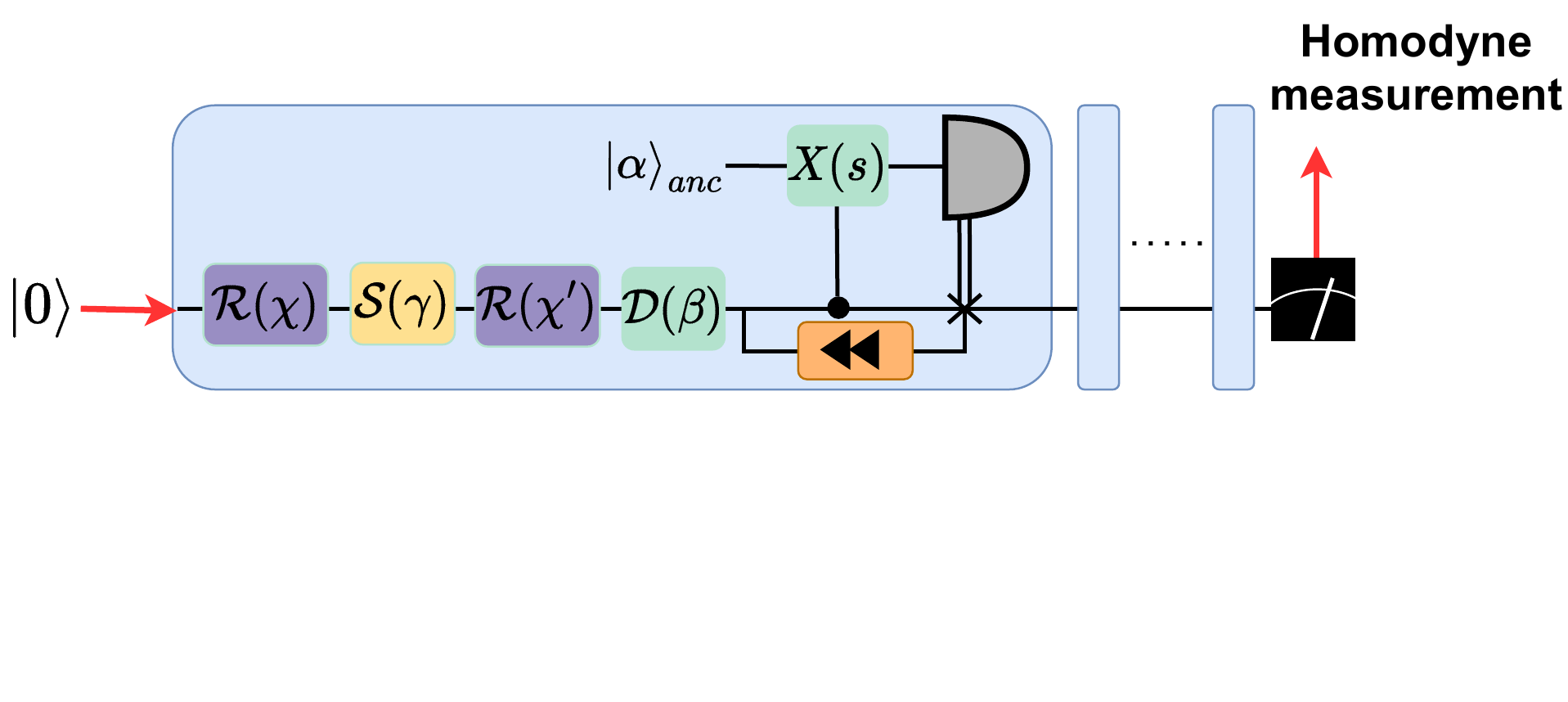}
    \caption{CV quantum NN architecture for quantum state preparation. The input to the network is the vacuum state, and at the end, a homodyne measurement is performed.}\label{fig:stateprep_arch}
\end{figure} 
For simplicity, we fixed the input state to be the vacuum, $\ket{\Psi_0}  = \ket{0}$. We considered a basic single-mode architecture of a quantum neural network with a fixed number of layers, as shown in Fig.~\ref{fig:stateprep_arch}. As described earlier, our goal is to find the parameters $\vec{\zeta}$ such that $U(\vec{\zeta}) \ket{ 0 } = \ket{ \Psi_t }$, or $| \bra{ \Psi_t } U(\vec{\zeta}) \ket{0} | = 1$.
For this case, we performed optimization by minimizing the cost function:
\begin{equation}
    C(\vec{\zeta}) = \left| \left|\bra{\Psi_t} U(\vec{\zeta}) \ket{0} \right|^2 -1 \right|~.
\end{equation}
To obtain $C$, we perform homodyne tomography on the final state $U(\vec{\zeta}) \ket{0}$. We measure the quadrature $X_\phi = \frac{1}{2} (e^{i\phi} a^\dagger + e^{-i\phi} a)$, and obtain a series of output pairs $(\phi_k, x_k)$ ($k=1,\dots, N$). They allow us to obtain an estimate of the Wigner function of the final state, $W_{U(\zeta)\ket{0}} (x,p)$. By comparing with the Wigner function of the desired state, $W_{\ket{\Psi_t}} (x,p)$, we deduce the cost function $C$ from the overlap
\be \left|\bra{\Psi_t} U(\vec{\zeta}) \ket{0} \right|^2 = 2\pi \int dx dp \, W_{U(\zeta)\ket{0}} (x,p) W_{\ket{\Psi_t}} (x,p)~. \ee

Although we focused on minimizing cost in the training process, we also calculated the fidelity between the target and optimized state. The fidelity measures how closely the optimized state matches the target state. It serves as another performance metric by providing another measure of accuracy with cost.

To test the performance of the quantum neural network, we prepared two different states, the single-photon state $\ket{\Psi_t} = \ket{1} = a^\dagger \ket{0}$, and the cat state
\be\label{eq:cat} \ket{\text{cat}, \theta} = \frac{1}{\sqrt{2}} \left( \mathcal{D} (\alpha_0) + e^{i\theta} \mathcal{D} (-\alpha_0) \right) \ket{0} \ee
where $\theta, \alpha_0 \in \mathbb{R}$.


Since we were using Strawberry Fields software, we could quickly get the Wigner function after performing the homodyne measurement. However, these Wigner functions are incompatible with the commonly used efficient TensorFlow framework to perform the optimization. Hence, we used the Python library - Scipy \cite{2020SciPy} to optimize cost as Scipy provided the Nelder-Mead optimization technique \cite{gao2012implementing}. This technique is non-gradient based and designed for high-dimensional minimization, which worked best for our purposes as our goal was to minimize the cost built from 2D Wigner functions obtained via homodyne detection.

\begin{figure}
\centering
    \includegraphics[scale=0.5]{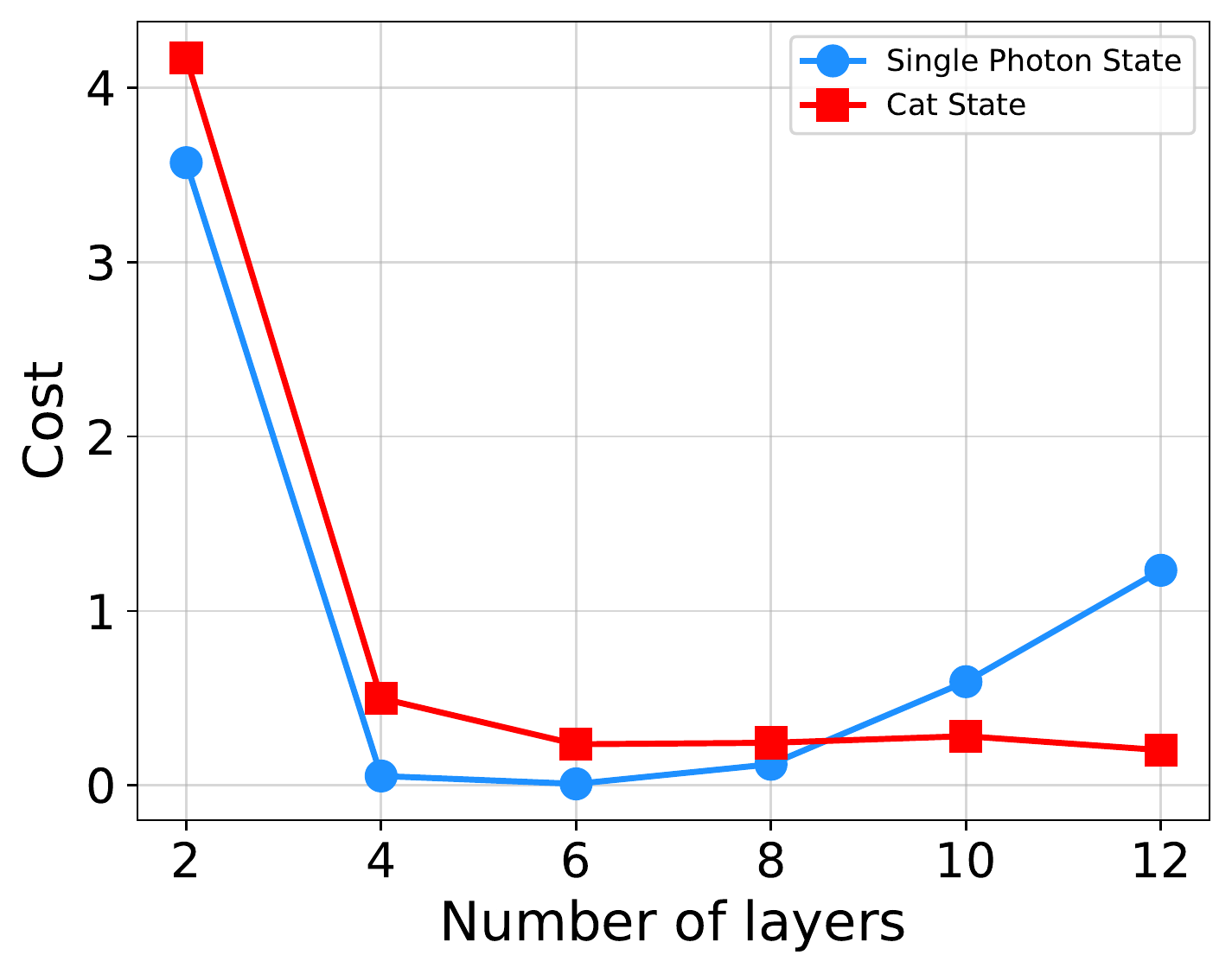}
    \caption{Value of cost function for different number of layers for preparation of single-photon (blue circles) and cat (red squares) states. Values averaged over 5 independent runs. Cutoff dimension was set to 6 (10) for the single-photon (cat) state.}\label{fig:layers}
\end{figure} 

We noticed that just in a few steps, the model started to learn the state. For best results, we ran the model with a different number of layers, as shown in Fig.~\ref{fig:layers}. In these simulations, we did not include any possible quantum hardware errors. A small number of layers yields a higher cost, but fewer layers require fewer gates and lead to fewer errors due to quantum hardware imperfections. Ignoring such errors, as the number of layers increases, the cost is lowered. Notice that the cost starts to increase again beyond a certain number of layers due to overfitting, e.g., for more than 12 layers in the case of the single-photon state. There is an optimum number of layers which would be important to determine by including a realistic model of quantum hardware. 

\begin{figure}
     \centering
     \begin{subfigure}{0.4 \textwidth}
         \centering
         \includegraphics[scale=0.28,trim={1.5cm 0.5cm 0 0cm}]{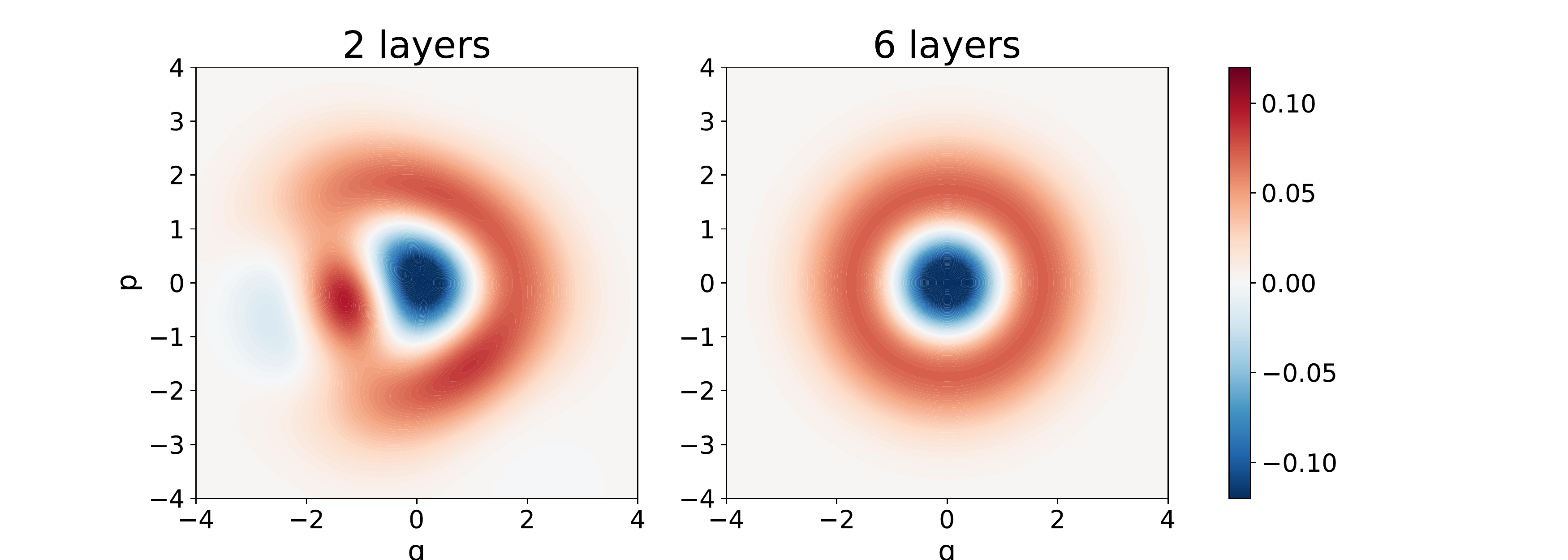}
         \caption{2D Wigner function comparison for 2 and 6 quantum layers, with fidelities 78.3\% and 99.9\%, respectively.}
     \end{subfigure}
     \hspace{-0.5em}
     \begin{subfigure}{0.4 \textwidth}
         \centering
         \includegraphics[scale=0.2]{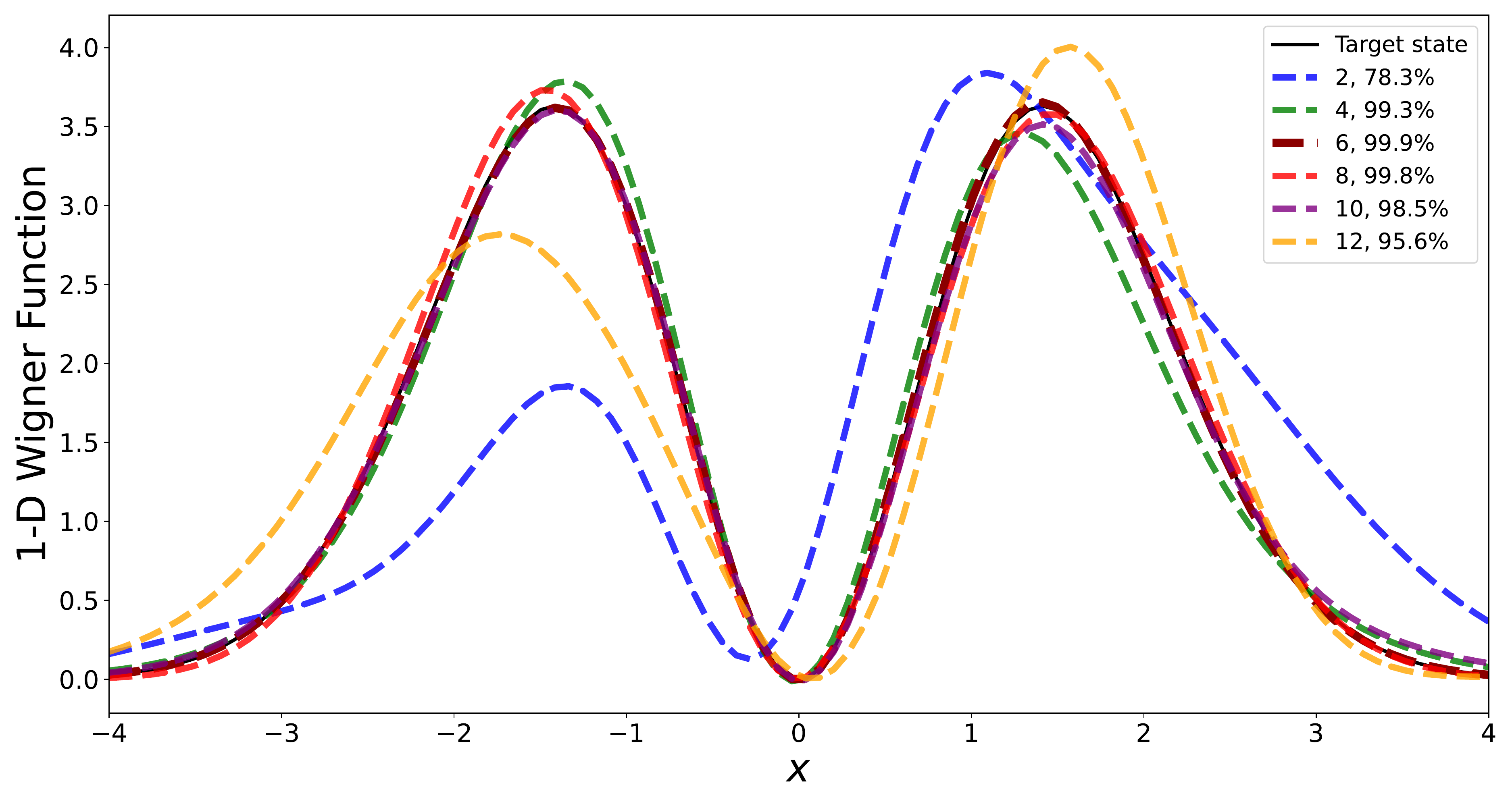}
         \caption{1D Wigner function for various numbers of quantum layers. 6 quantum layers produce the best fidelity. }
     \end{subfigure}
    \caption{Study of the effect of changing the number of layers of the QNN for the preparation of a single-photon state. The maximum number of steps was kept fixed at 5000 and the cutoff dimension at 6.}\label{fig:SPsinglephoton}
\end{figure}

For the single photon state, 6 quantum layers gave the best results with a fidelity of 99.9\%. The cost achieved after optimization was 0.008 after 5000 steps. We used a cutoff dimension of 6, and the maximum number of steps was fixed at 5000. It should be noted that in the Nelder-Mead optimization, the number of steps is determined by the difference between two consecutive cost values. This difference is treated as another hyperparameter during the training process. The result of comparing other numbers of layers is shown in Fig.~\ref{fig:SPsinglephoton}. We plotted the 2D Wigner function for 2 and 6 layers to demonstrate the importance of the optimal number of layers. We also showed the 1D Wigner function for different quantum layers. We integrated the 2D Wigner function over the momentum using the Scipy library to obtain the 1D Wigner function. The plot shows that the results improve up to 6 quantum layers, and beyond that, they worsen due to overfitting.

\begin{figure}[ht!]
    \includegraphics[scale=0.28,trim={1.5cm 0.5cm 0 0cm},clip]{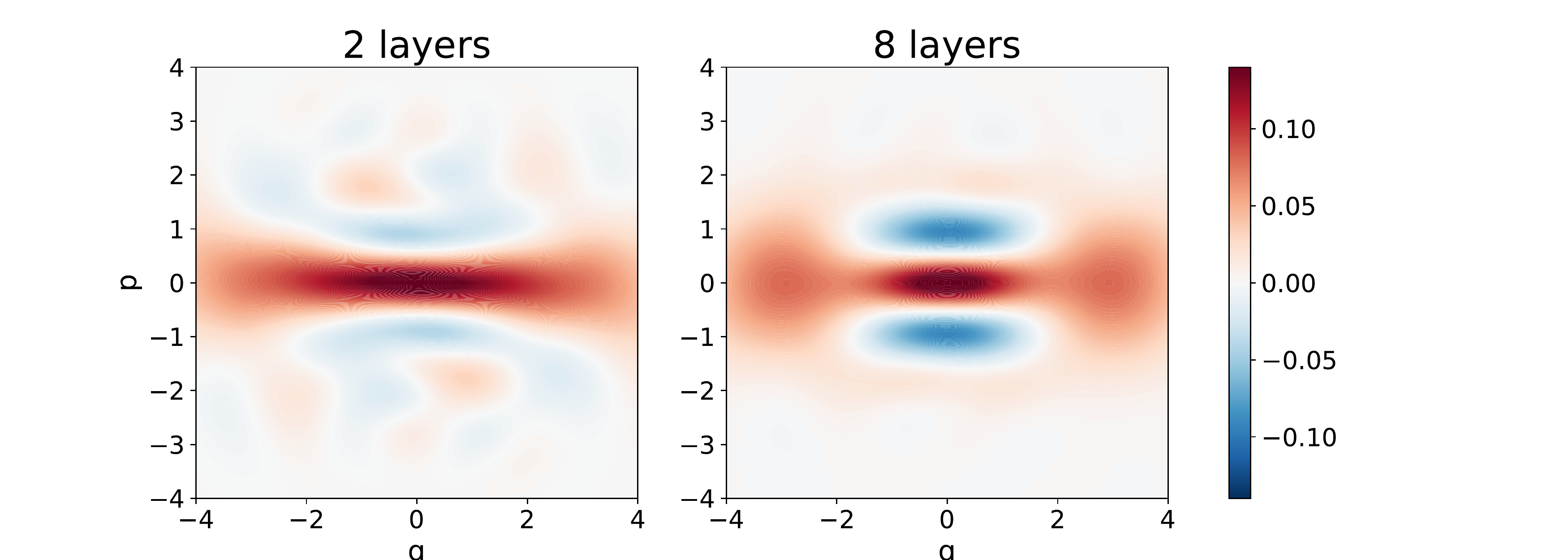}
    \caption{Preparation of an even cat state (Eq.\ \eqref{eq:cat} with $\alpha_0 = 1.5$, $\theta = 0$). 2D Wigner function comparison for 2 and 8 quantum layers with fidelity 79\%  and 99.8\%, respectively. A cutoff dimension of 10 was used in both cases. 2 (8) layers optimized in 2000 (9800) steps resulting in a cost of 2.64 (0.03).}\label{fig:SPcat}
\end{figure}

Turning to state preparation of the cat state \eqref{eq:cat}, which is a superposition of two coherent states, we concentrated on the even cat state with $\alpha_0 =1.5$ and $\theta =0$. As the cat state is more complicated than the single photon state, we had to increase the cutoff dimension to 10. We achieved high fidelity with 8 quantum layers. The comparison of the 2D Wigner functions of 2 and 8 layers is shown in Fig.~\ref{fig:SPcat}. We obtained excellent fidelity of 99.8\% with 8 layers, and after training, the cost was minimized to the value of 0.03 after 9800 steps. For comparison, with 2 layers, we obtained fidelity of 79\% with cost at 2.64 after 2000 steps.

\begin{figure}[ht!]
    \includegraphics[scale=0.3,trim={0.8cm 2.8cm 0 0cm},clip]{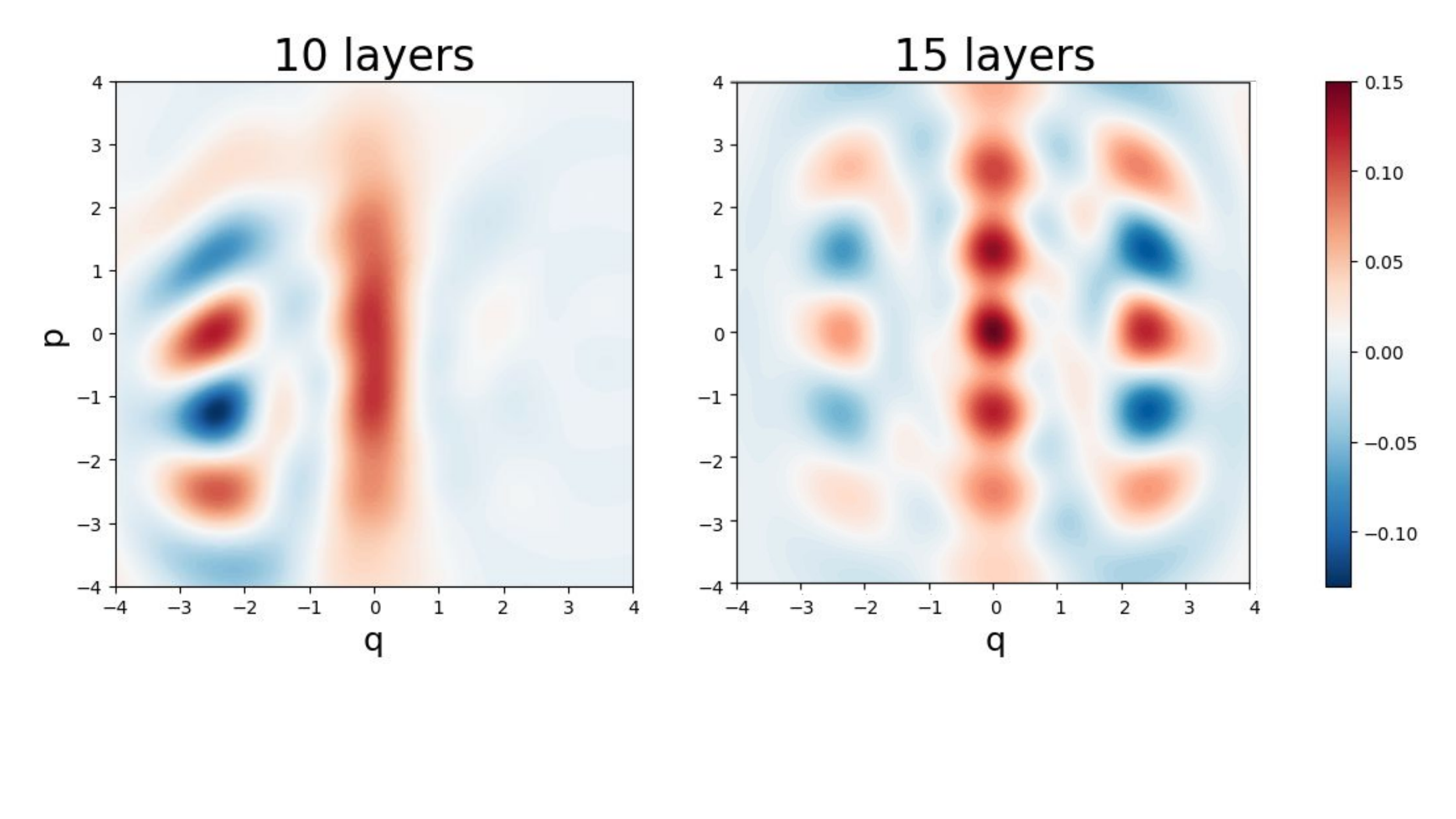}
    \caption{Preparation of a realistic GKP state (Eq.\ \eqref{eq:gkp} with $\epsilon = 0.1$). 2D Wigner function comparison for 10 and 15 quantum layers with fidelity 70.6\%  and 93.9\%, respectively. A cutoff dimension of 15 was used in both cases. 10 (15) layers were optimized in 18000 (15000) steps resulting in a cost of 4.87 (1.1).}
    \label{fig:SPgkp}
\end{figure}

We also prepared a realistic GKP state \cite{PhysRevA.64.012310}. Ideal GKP states are linear combinations of an infinite number of eigenstates of the $q$-quadrature. We concentrated on the state
\be \ket{\mathrm{GKP}}_{\mathrm{ideal}} = \sum_{n=-\infty}^\infty \ket{q= 2n\sqrt{\pi \hbar}}  \ee
However, such states are not normalizable and impossible to create experimentally because they have infinite energy and each component would require an infinite amount of squeezing. For a realistic case, we applied an energy cutoff and defined the realistic GKP state \cite{shi2019fault}
\be\label{eq:gkp} \ket{\mathrm{GKP}}_{\mathrm{real}} = e^{-\epsilon a^\dagger a} \ket{\mathrm{GKP}}_{\mathrm{ideal}} \ee
We chose $\epsilon=0.1$. Since this state is more complex than a cat state, we had to increase the cutoff dimension even further to 15 and employ 15 layers. After 15000 steps, we achieved a fidelity of 93.9\% at a cost of 1.1. The comparison between 10 and 15 layers is shown in Fig.~\ref{fig:SPgkp}. 


\subsection{Curve Fitting} \label{sec:IIIB}
Next, we build a CV QNN in a supervised learning setting to learn the relationship between input ($x$) and output ($f(x)$), also known as curve fitting. It is an essential part of data analysis and a classic machine-learning problem. 

\begin{figure}[ht!]
    \centering
    \includegraphics[scale=0.4, trim={3cm 3cm 4cm 0cm}]{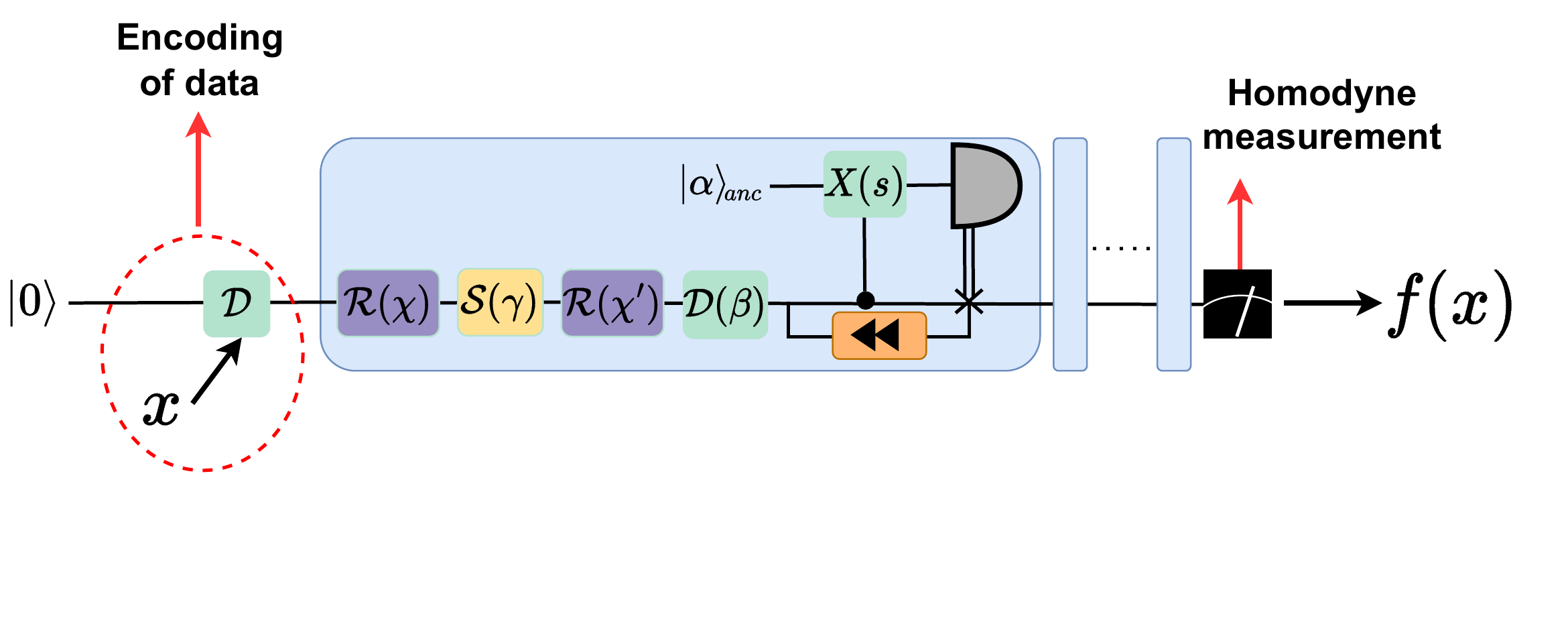}
    \caption{CV quantum NN architecture for curve fitting. }
    \label{fig:arch_curvefit}
\end{figure}
\begin{figure}[ht!]
     \centering
     \begin{subfigure}[b]{0.23 \textwidth}
         \centering
         \includegraphics[width=\textwidth]{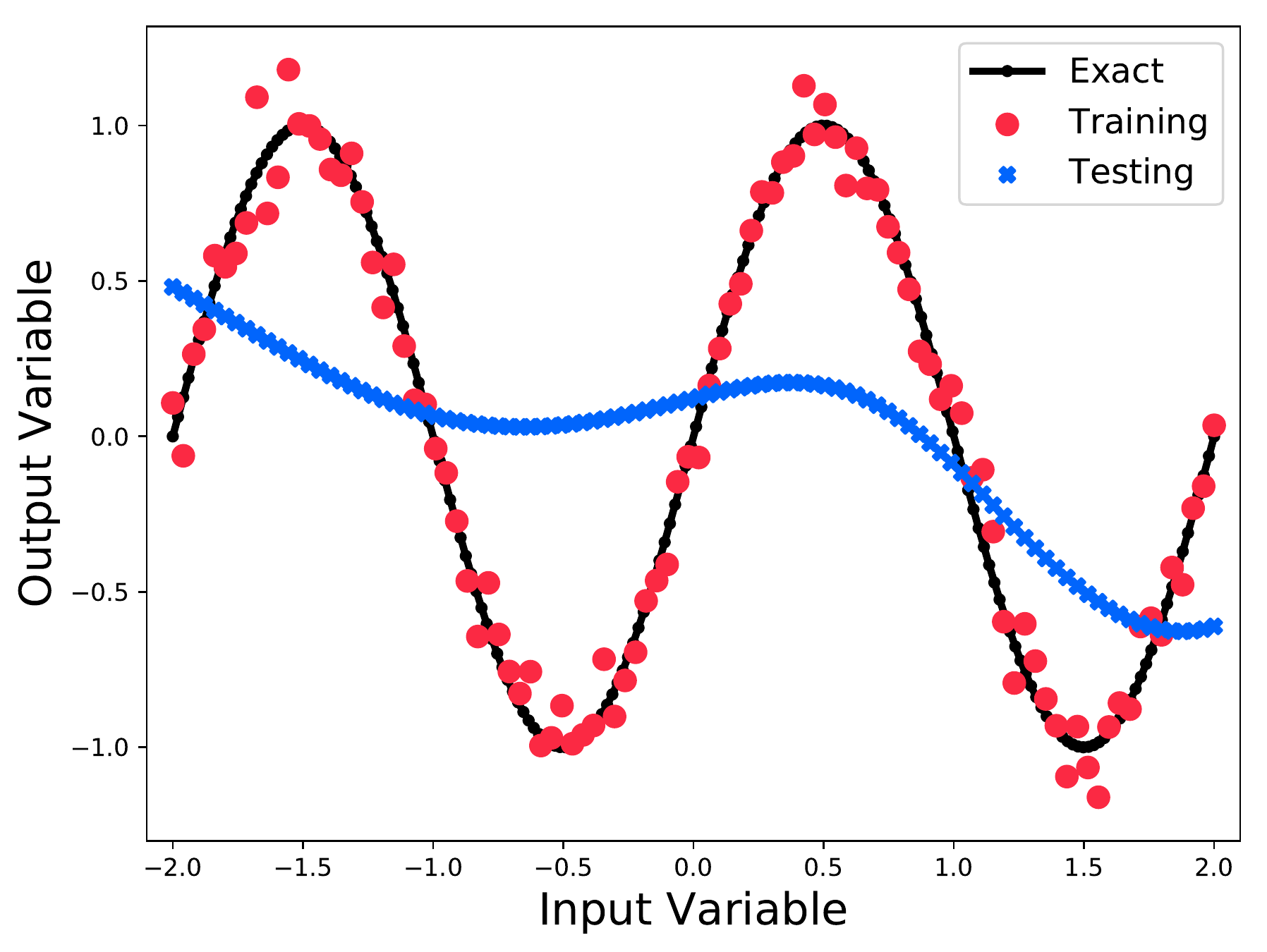}
         \caption{2 layers, cost $=0.318$}
     \end{subfigure}
     \hspace{-0.5em}
     \begin{subfigure}[b]{0.23 \textwidth}
         \centering
         \includegraphics[width=\textwidth]{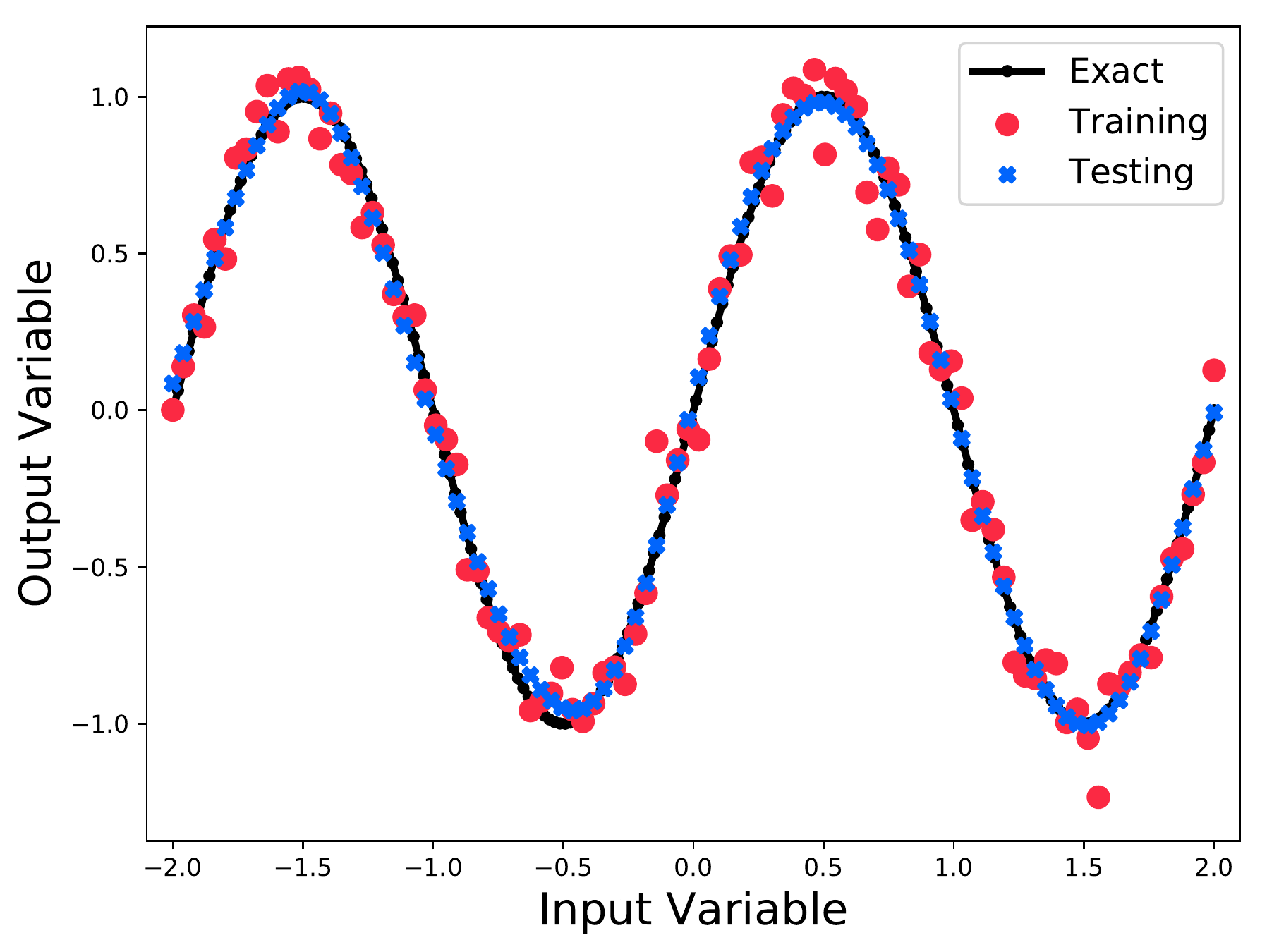}
         \caption{6 layers, cost $=0.008$}
     \end{subfigure}
        \caption{The effect of changing the number of layers on curve fitting of a noisy sine function. 100 data points were used for training and testing in 1000 steps with error parameter $\epsilon=0.1$ and cutoff dimension 6.}
        \label{fig:CF_layers}
\end{figure}
\begin{figure}[ht!]
    \centering
    \includegraphics[scale=0.4]{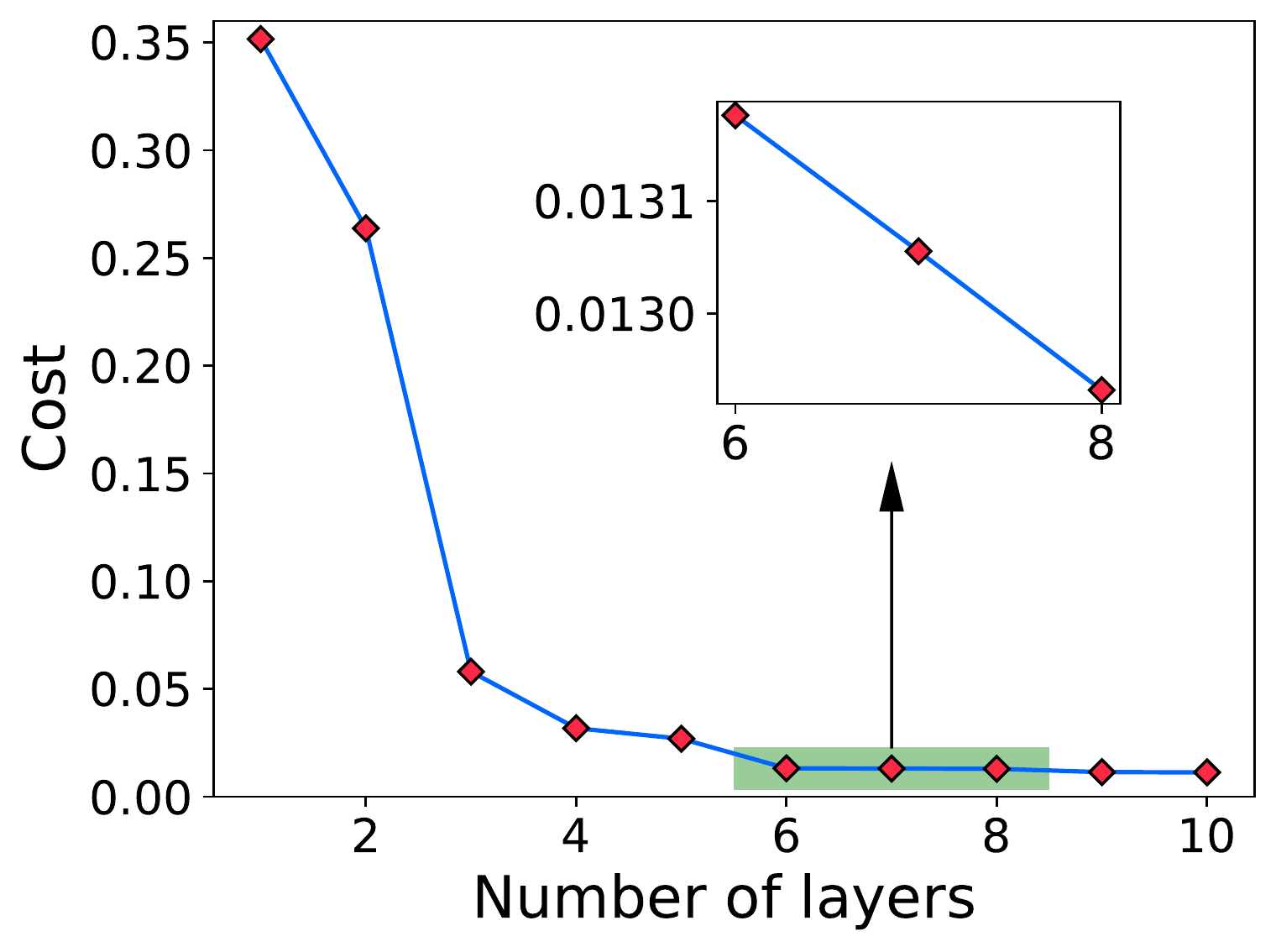}
    \caption{Cost function values vs. the number of layers plot obtained after performing learning on our CV quantum neural network model for fitting the noisy sine function curve. Each data point in the scatter plot corresponds to the $5$ independent runs average. The training was done for 1000 steps with 100 data points (with $\epsilon=0.1$) and a cutoff dimension of 6.}
    \label{fig:costvslayers_CF}
\end{figure}

The architecture for our CV QNN is shown in Fig.~\ref{fig:arch_curvefit}.
We encoded the classical input, $x$, sampled from a noisy function, $f(x)$, as the coherence parameter of the input qumode, $\ket{x} = \mathcal{D} (x) \ket{0}$.
The objective was to train the CV QNN to generate output states $\ket{\psi_x}$ that have an expectation value of the quadrature, $q$, close to $f(x)$ (i.e., $\bra{\psi_x} q \ket{\psi_x} \approx f(x)$) for a given input $x$. We studied the noisy sine function. The data were prepared as $f(x) = f_0(x) + \Delta f$ where $\Delta f$ is a normal distribution with zero mean and standard deviation $\epsilon$. The parameter $\epsilon$ determines the amount of error present in the training data. We chose the noisy sine function with $f_0(x) = \sin{x}$ in the range of $x \in(-2,2)$. We used 6 quantum layers in this process. The training was done on 1000 steps with a Hilbert-space cutoff dimension of 6. The training and test data were prepared as tuples $(x_i, f(x_i))$, and $x_i$ was chosen uniformly at random in the chosen interval. For training, we chose the cost function to be the mean square error (MSE) value between the circuit outputs and the desired function values,
\begin{equation}
    C = \frac{1}{N} \sum_{i=1}^N \left [ f(x_i) - \bra{\psi_{x_i}} q \ket{ \psi_{x_i}}  \right ]^2~.
\end{equation}
To learn about the performance of CV QNN, we studied how the cost would change if we increased the number of layers. This helped us determine the optimum number of layers required for the desired results. We started with a single quantum layer and increased the number of quantum layers up to 10. We kept the number of steps fixed at 1000 with 100 data points and a cutoff dimension of $6$. We used the Adam optimizer to minimize the cost function value. Some interesting results of how testing data behaved with a changing  number of layers are shown in Fig.~\ref{fig:CF_layers}. The final result of the study is summarized in the cost vs.\ the number of layers in Fig.~\ref{fig:costvslayers_CF}. We found that the cost function value decreases as we increase the number of layers, making the curve fit better. However, this improvement saturates around 6 quantum layers. One also has to keep in mind that more layers correspond to more number of training parameters. Hence, finding a number that yields good results and keeps the training parameters manageable is important. In the case considered here, optimal results were obtained with 6 layers. 

\begin{figure}[ht!]
     \centering
     \begin{subfigure}[b]{0.23 \textwidth}
         \centering
         \includegraphics[width=\textwidth]{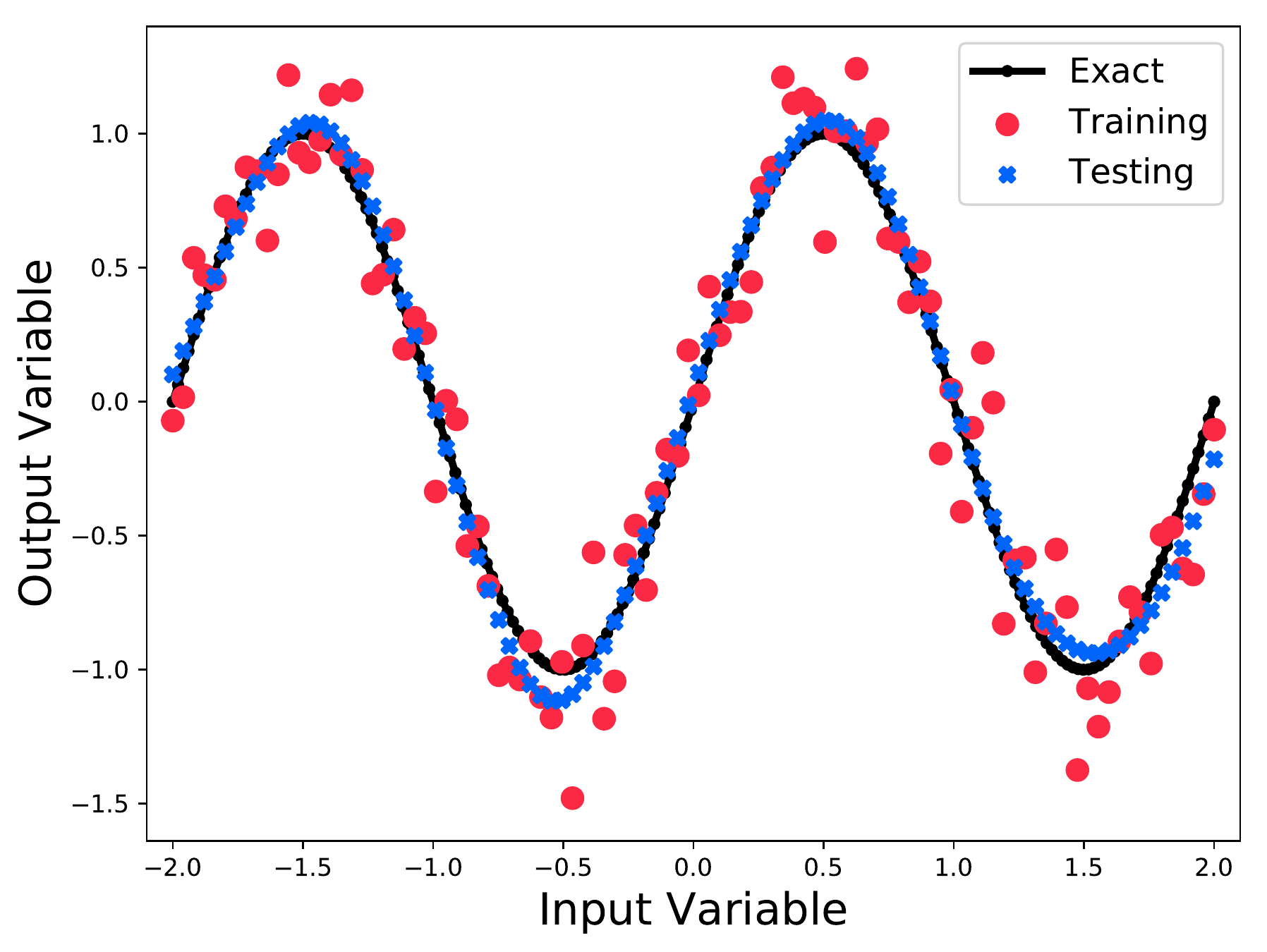}
         \caption{$\epsilon = 0.2$ and cost$=0.037$}
     \end{subfigure}
     \hspace{-0.5em}
     \begin{subfigure}[b]{0.23 \textwidth}
         \centering
         \includegraphics[width=\textwidth]{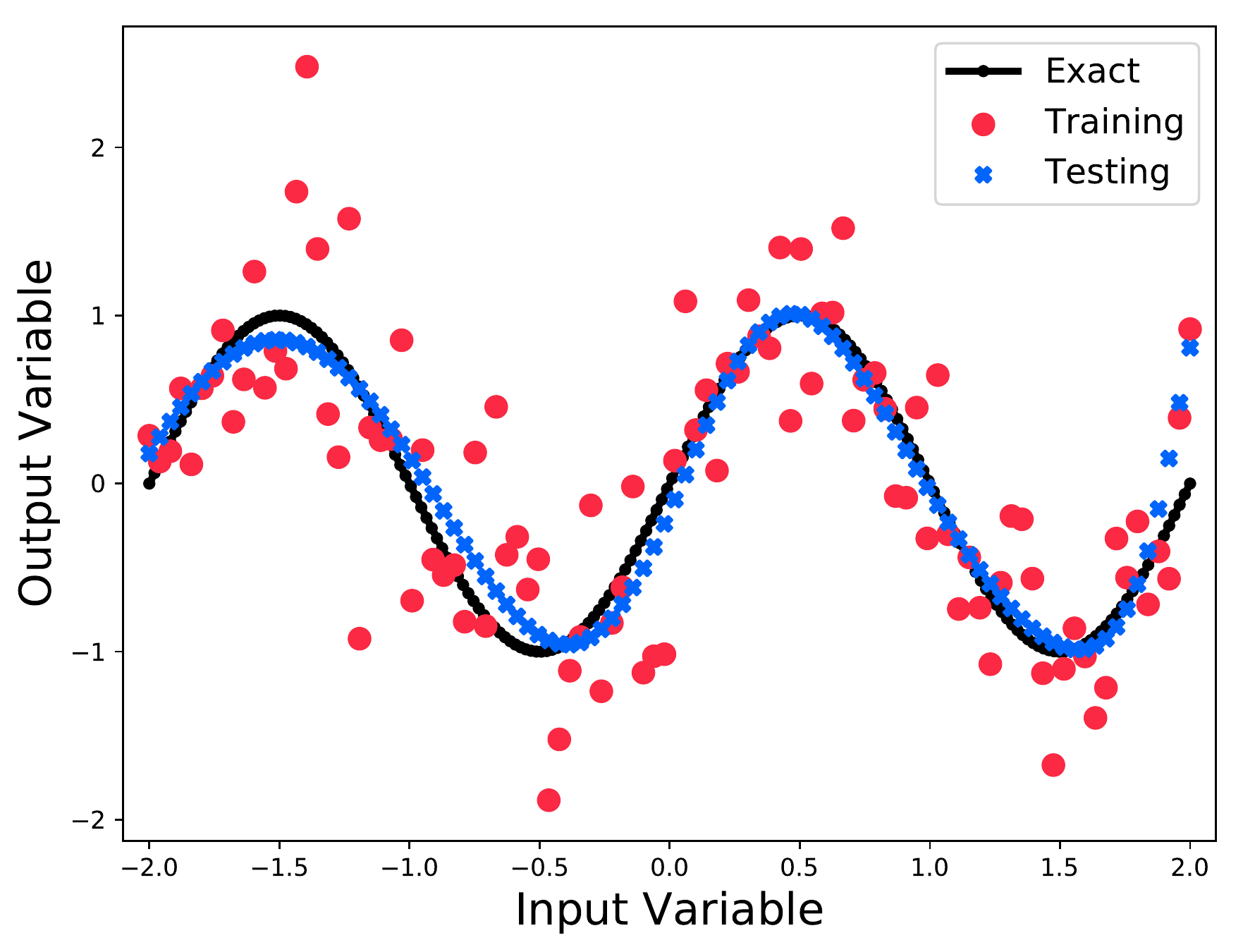}
         \caption{$\epsilon = 0.5$ and cost$=0.232$}
     \end{subfigure}
        \caption{The effect of increasing the error in the data on  curve fitting of a noisy sine function. 100 data points were used for training and testing in 1000 steps with cutoff dimension 6 using 6 quantum layers.}
        \label{fig:CF_error}
\end{figure}

\begin{figure}[ht!]
    \centering
    \includegraphics[scale=0.4]{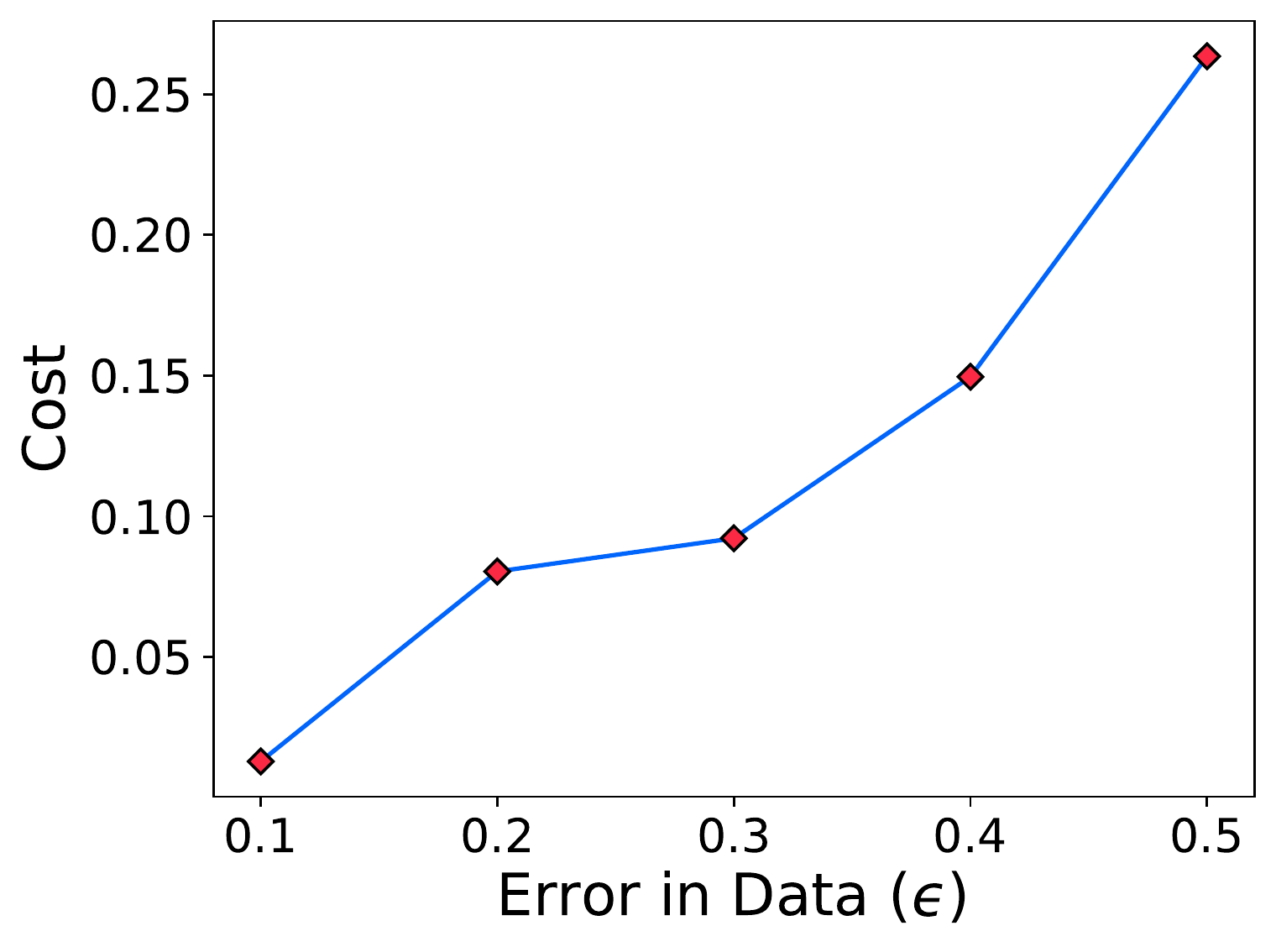}
    \caption{Plot of cost function value vs.\ noise in data determined by the parameter $\epsilon$, obtained after learning on a QNN to fit the noisy sine function curve. Each data point in the scatter corresponds to the average of 5 independent runs. The training was done for 1000 steps with 100 data points, a cutoff dimension of 6, and using 6 quantum layers. }
    \label{fig:costvserror}
\end{figure}

We also studied how the noise present in data affected our results by varying the error parameter $\epsilon$ discussed above. We kept the number of steps fixed at 1000 with 100 data points, a cutoff dimension of 6, and 6 quantum layers. The results of $\epsilon=0.2, 0.5$ are shown in Fig.~\ref{fig:CF_error}. The value of cost increases from 0.037 to 0.232 as we increase the error from $\epsilon = 0.2$ to $\epsilon=0.5$, and the fitting worsens as we increase the noise; this is expected as the model is training on noisy data. Although the fitting is getting worse with increased noise, the CV QNN still performs well in learning the shape of the sine function. The complete study of the dependence of the cost function on data noise ($\epsilon$) is shown in Fig.~\ref{fig:costvserror}. 

\subsection{Binary classification} \label{sec:IIIC}
For the third problem, we constructed a CV quantum-classical hybrid neural network to demonstrate its effectiveness in detecting fraudulent transactions on credit card purchase data. This is a binary classification problem, a canonical problem in machine learning. The main reason for including the classical layers in this problem is that we want to encode the data with the help of classical layers. 

The credit card transaction data is taken from Kaggle \cite{kaggle}, a publicly available database. Each transaction was flagged as either genuine or fraudulent with 28 features. Only 0.172\% of transactions were fraudulent out of a total number of 284,807. 

\begin{figure}[ht]
\centering
\includegraphics[scale=0.4]{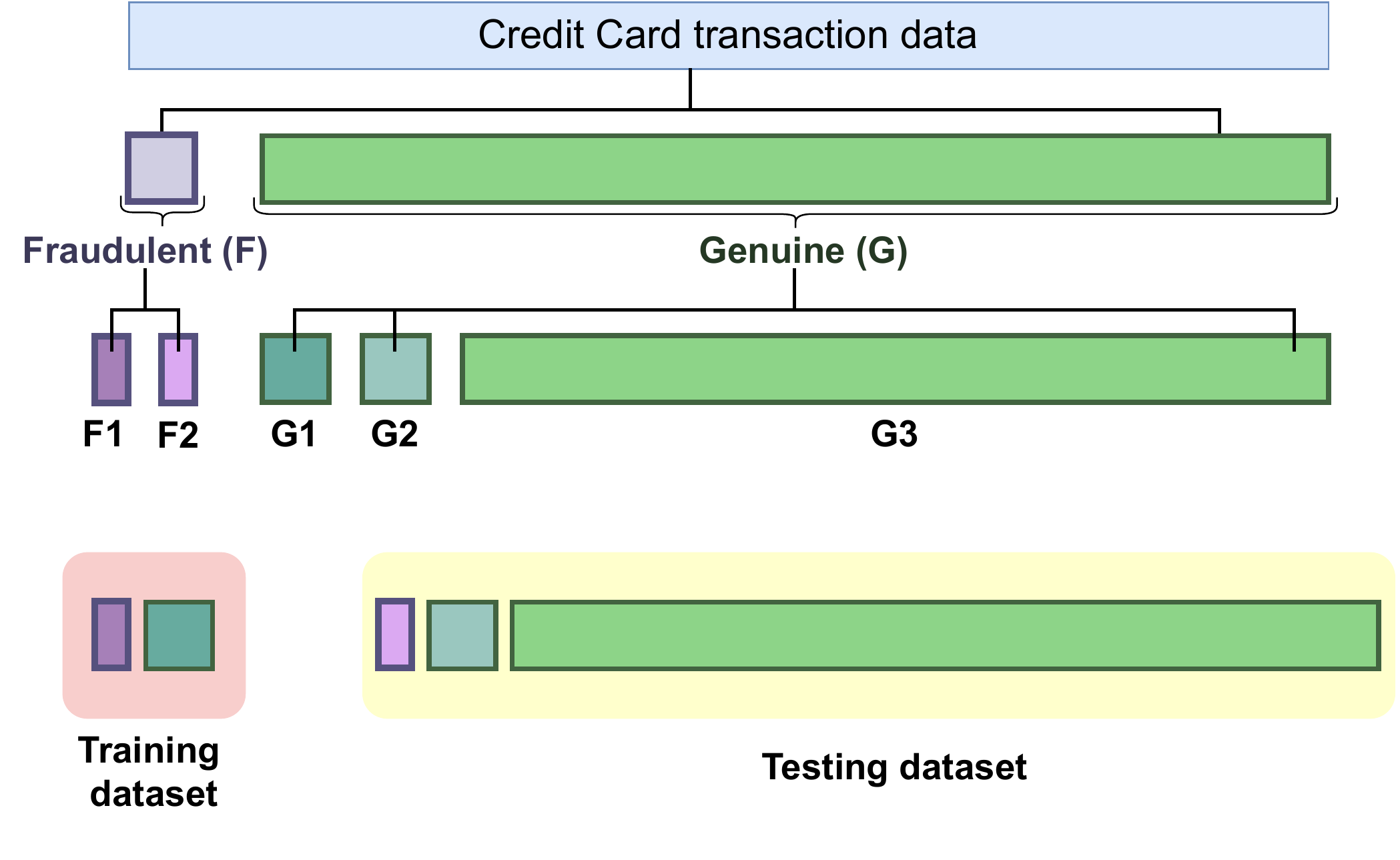}
\caption{The description of data preparation done for binary classification problem. Note that the smaller parts of fraudulent and genuine transactions are denoted by the letters F1 and F2 and G1 and G2, respectively. The size of G1 and G2 are equal but three times more than F1 and F2 because of 3:1 undersampling. In the end, the training dataset consists of F1 and G1, and the testing dataset consists of F2, G2, and G3.}
\label{fig:dataprep}
\end{figure}

First, we split the data into training and testing parts. In the training data set, we under-sampled the genuine transactions  by selecting them randomly and ensuring that the genuine-to-fraudulent transaction ratio was 3:1. All the remaining genuine transactions were added to the test data set. This data preparation is explained in detail in Fig.~\ref{fig:dataprep}.   

\begin{figure*}
\centering
\includegraphics[width=0.7 \textwidth,  trim={4cm 5cm 4cm 0cm}]{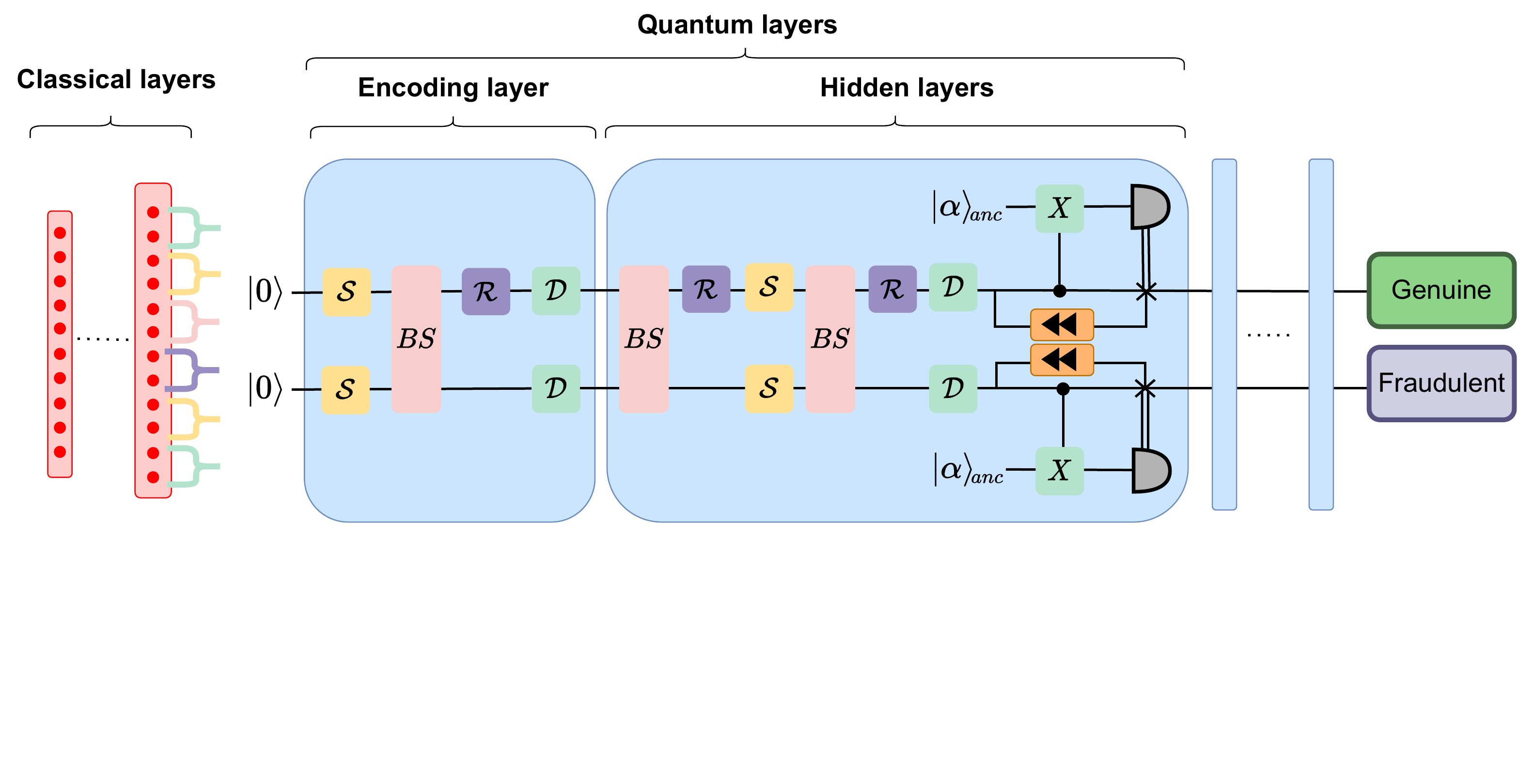}
\caption{CV hybrid NN for fraudulent transaction detection in credit card data. The parameters of each gate in the encoding layer are obtained from the values of the last classical layer. }
\label{fig:architecture}
\end{figure*}

The network architecture is shown in Fig.~\ref{fig:architecture}. Four fully connected feed-forward classical layers are followed by five quantum layers with four modes (two are ancillary qumodes). The credit card data is fed into the first classical layer of size 10, followed by two hidden layers of the same size. The last classical layer of size 12 controls the gate parameters in the first quantum input layer. This layer marks the beginning of the quantum part of the neural network. And because we are letting the last classical layer control the gate parameters, the quantum layer is the encoding layer. We start with four vacuum qumodes. These layers contain two single-mode squeezing gates, $S$, one interferometer gate, $U$, two displacement gates, $D$, and two $\textup{CX}$ gates, which provide non-linearity through measurement on the ancilla qumode. At the end of the encoding quantum layer, a photon number measurement is performed on the two ancillary qumodes. The two primary qumodes are allowed to advance when the detectors on the ancillary qumodes click. If they do not click, the main qumodes are fed back into the quantum circuit, as shown in Fig.\ \ref{fig:architecture}. The feedback loops are repeated until the detectors on the ancillary qumodes click, thus implementing the desired non-linearity in the CV quantum neural network \cite{marshall2015repeat}. We repeat this process for four more hidden layers. Finally, we measure the photon number on the two output primary qumodes that emerge after the last quantum output layer. If we find the photon in the first qumode, we call it a genuine transaction; if we find it in the second qumode, it is a fraudulent transaction.

The training was performed using the Adam Optimizer with a batch size 24. We minimized the cost function defined by: 
\begin{equation}\label{eq:loss_func}
    C = \Sigma_{i \in \textup{data}} \left (1-p_i \right)^2~,
\end{equation}
where $p_i$ is the probability of detecting a photon for input $i$ in the correct mode.
We used a cutoff dimension of 8 in each mode for 10,000 batches. Once the model was trained, we tested it by choosing a threshold probability closest to the optimal ROC, required for a transaction to be classified as genuine.  

\begin{figure}
     \centering
     \begin{subfigure}{0.5 \textwidth}
         \centering
         \includegraphics[scale=0.36,trim={0.0cm 0.0cm 0.0cm 0cm}]{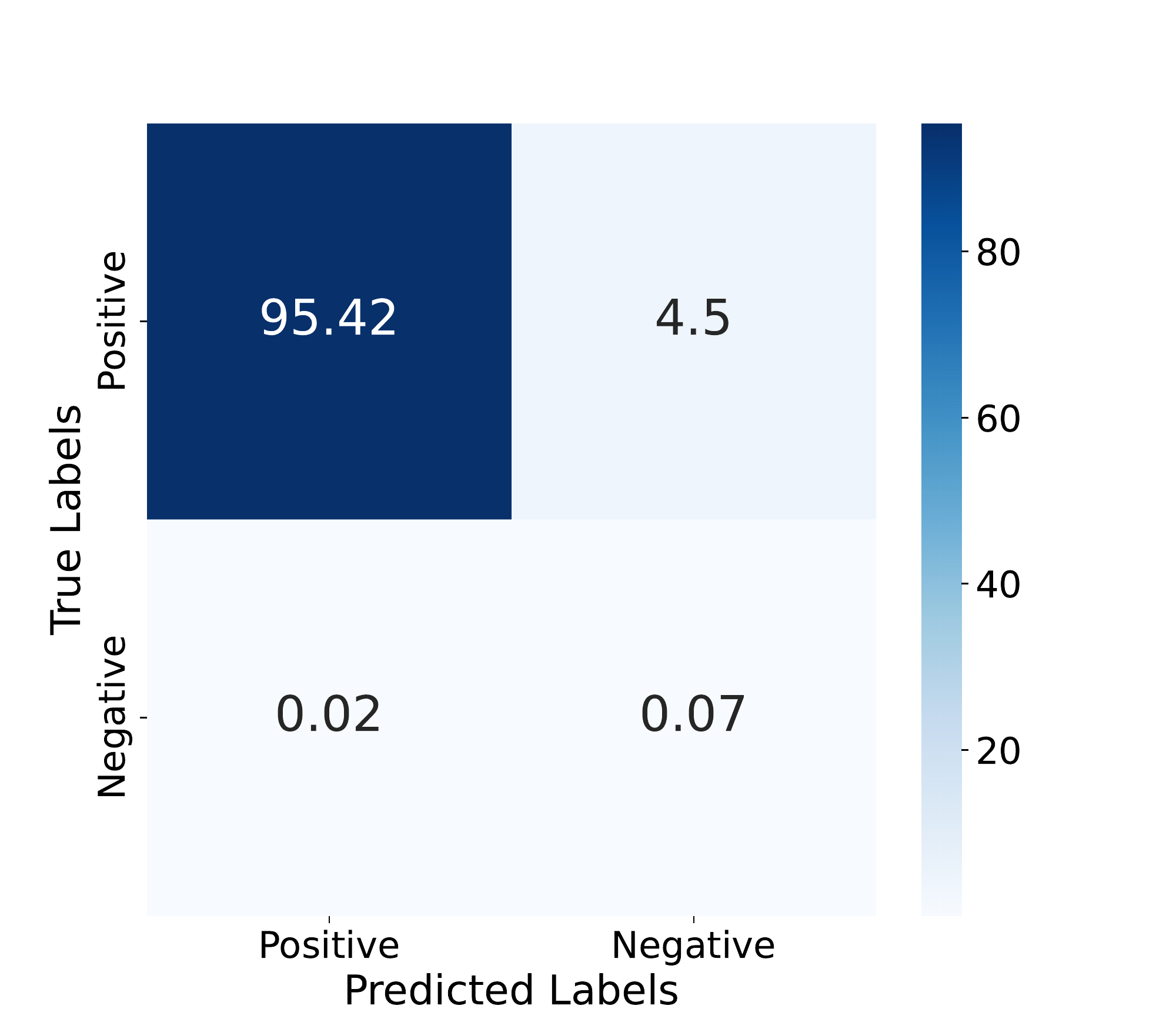}
         \caption{Confusion Matrix}
     \end{subfigure}
     \hspace{-0.5em}
     \begin{subfigure}{0.4 \textwidth}
         \centering
         \includegraphics[scale=0.5]{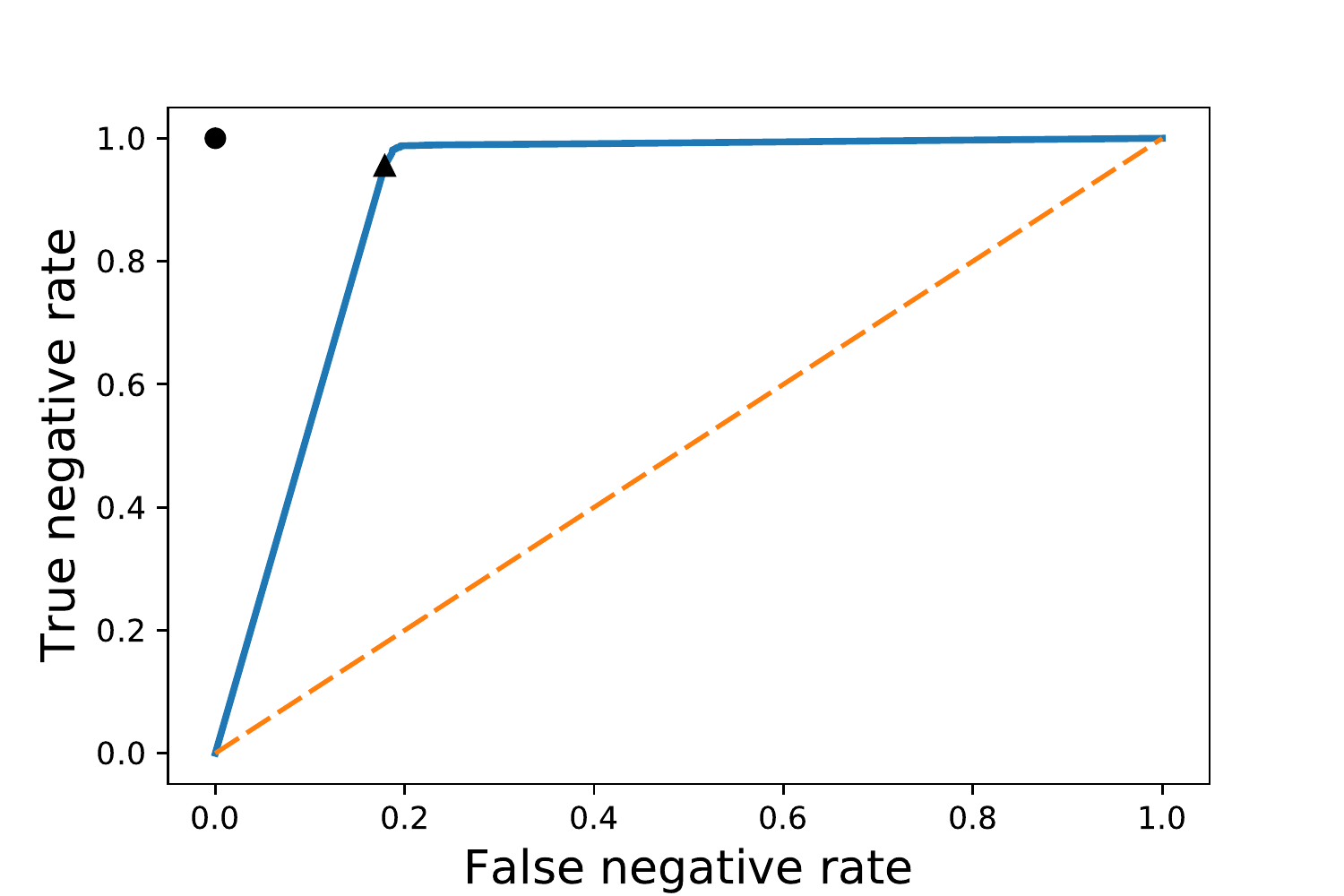}
         \caption{ROC curve}
     \end{subfigure}
    \caption{Results of fraud detection in credit card transactions. The CV hybrid NN had 2 classical and 4 quantum hidden layers. The accuracy from the confusion matrix is 95.48\%, and the area under the ROC curve is 0.9.}\label{fig:result2}
\end{figure}


The confusion matrix and Receiver Operating Characteristic (ROC) curve are shown in Fig.~\ref{fig:result2}. The accuracy of the model calculated from the confusion matrix came out to be $95.48$. Also, the confusion matrix shows that the model predicts the genuine data correctly for more than 95\%. The number in the second quadrant representing the False Negative (FN) appears to be high. However, credit card companies can alert their users about such transactions, and by verifying these transactions, the FN can be brought down. The essential quadrant to consider is the third one that represents False Positive (FP), i.e., the fraudulent transactions that are wrongly predicted as genuine. Fortunately, this number is very low for the trained model. Also, the testing data set (which was used to plot the confusion matrix) contains a tiny number of fraudulent transactions, all identified almost correctly, as the percentage of fraudulent traction in the testing data set matches the fourth quadrant.
The circle dot in the ROC curve represents the ideal point, and the triangle is the closest point to the optimal within the chosen threshold. The area under the curve (AUC) is 0.90, which is close to the ideal value of 1. The AUC value is a good measure of the separability of the data being classified. 
\begin{figure}[ht!]
\centering
\includegraphics[scale = 0.35]{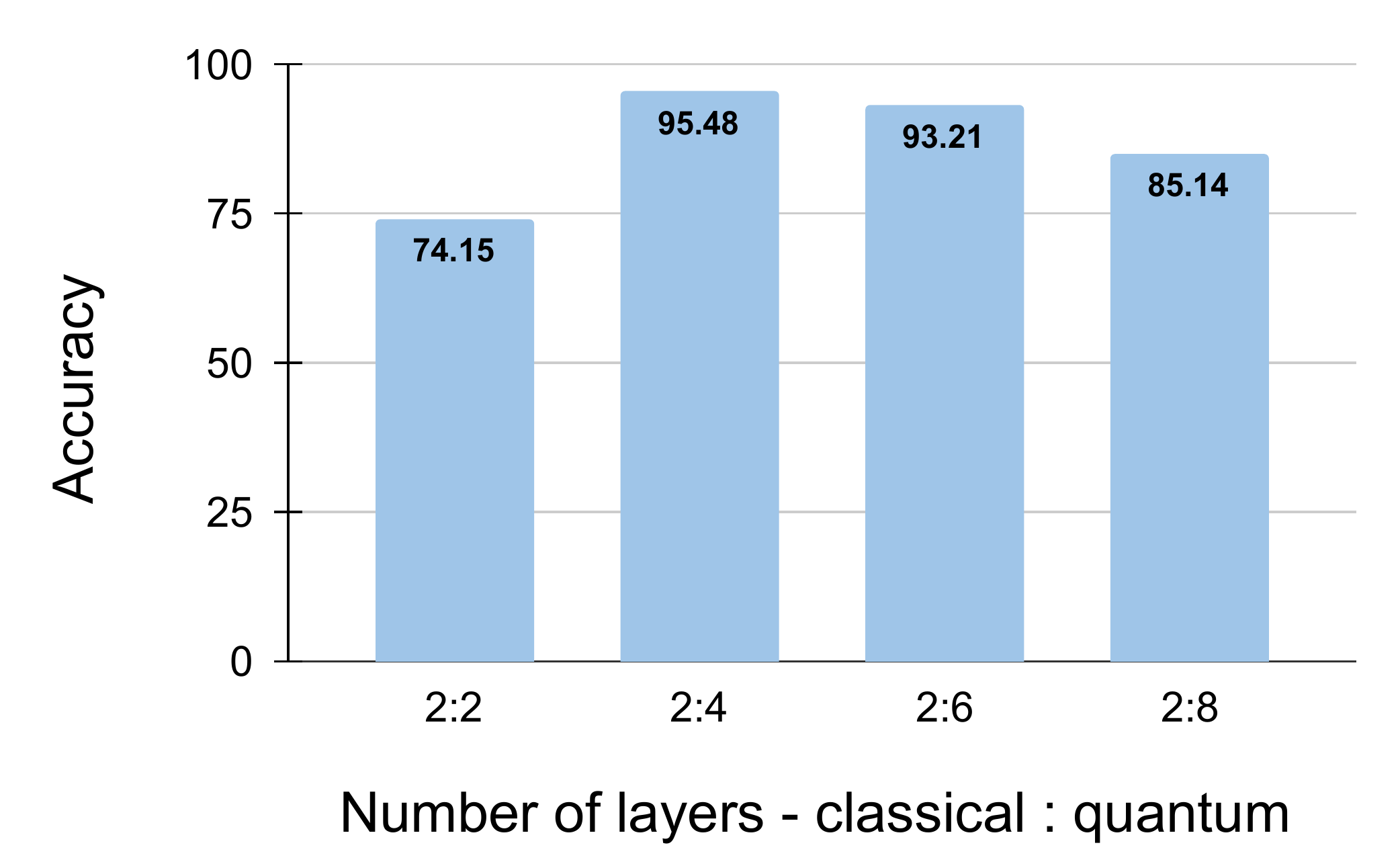}
\caption{Plot showing the effect of changing the number of classical and quantum layers in a hybrid CV NN on the accuracy of the results for binary classification.}\label{fig:diffqlayers}
\end{figure} 

It should be pointed out that the number of features of the credit card transaction data we are interested in is equal to the number of parameters in the quantum circuit used for data encoding. Therefore, we investigated the role of the classical layer in this hybrid quantum-classical neural network architecture. We kept the classical layers constant at 2 and changed the quantum layer at an increment of 2. The results are plotted as accuracy calculated from the confusion matrices vs. different layers in Fig.~\ref{fig:diffqlayers}. There is an optimal number of hybrid layer scenario which give out the best results. When we use 2 classical and 2 quantum layers, the accuracy is around 74\%, meaning there are not enough layers for learning. However, when we increased the quantum layers to 8 then, also the accuracy went slightly down, indicating the overtraining for a simple case of binary classification. We also understand that hyperparameters play a major role in training, but we found this for the set of hyperparameters we chose.

\subsection{Multi-label classification}
\label{sec:IIID}
Extending the results for binary classification, we developed a CV quantum-classical hybrid NN to classify MNIST  handwritten digits \cite{deng2012mnist} into their respective classes. The MNIST dataset comprises 60,000 training images and 10,000 testing images, each normalized to $28\times28$ pixels in size and grayscale in color. Each data point is labeled with the corresponding digit (0-9). Our current hardware limitations allowed our model to train and classify images up to 4 classes (0-3).

\begin{figure}[ht!]
\centering
\vspace{1cm}
\includegraphics[width=0.3 \textwidth,  trim={3cm 0cm 2cm 2.5cm}]{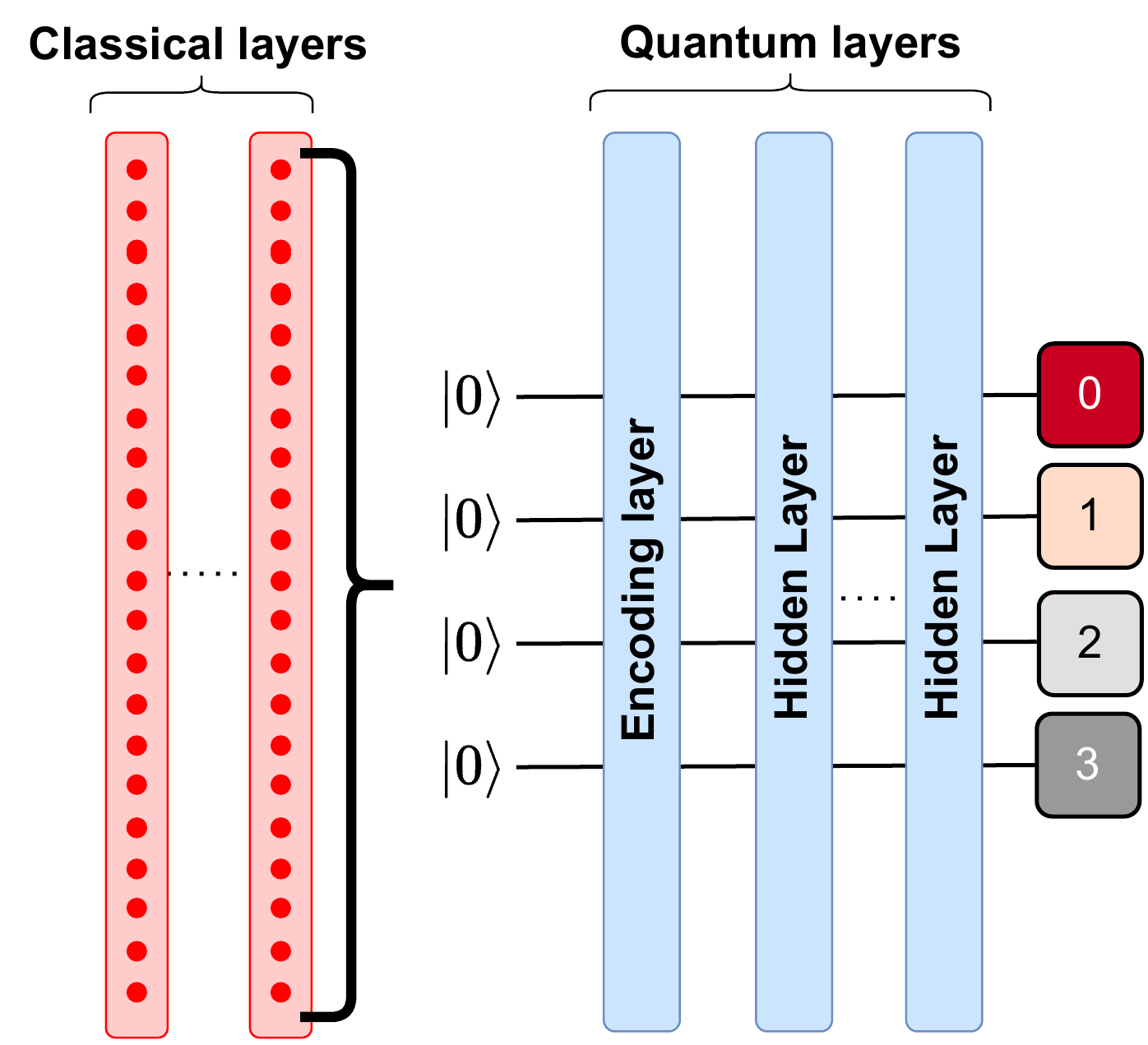}
\caption{CV hybrid NN architecture used for MNIST image classification. }\label{fig:mnist_arch}
\end{figure}

\begin{figure}[ht!]
\centering
\begin{subfigure}{0.4\textwidth}
    \includegraphics[width=0.4 \textwidth,  trim={1.5cm 0cm 1.5cm 0cm}]{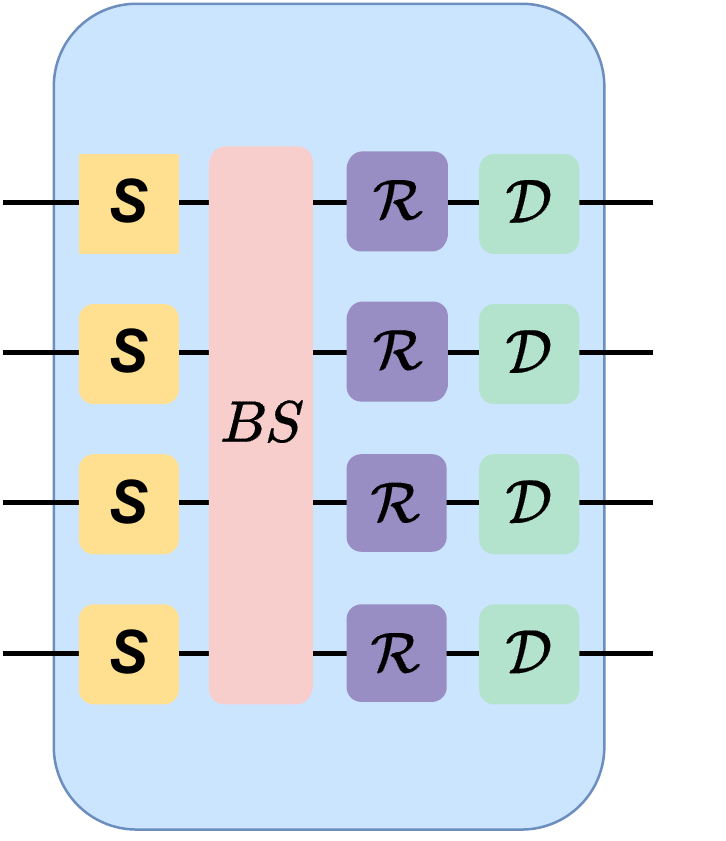}
    \caption{Encoding layer}\label{fig:mnist_enc}
\end{subfigure}
\begin{subfigure}{0.4\textwidth}
\vspace{1cm}
\includegraphics[width=1. \textwidth,  trim={0cm 6cm 0cm 2cm}]{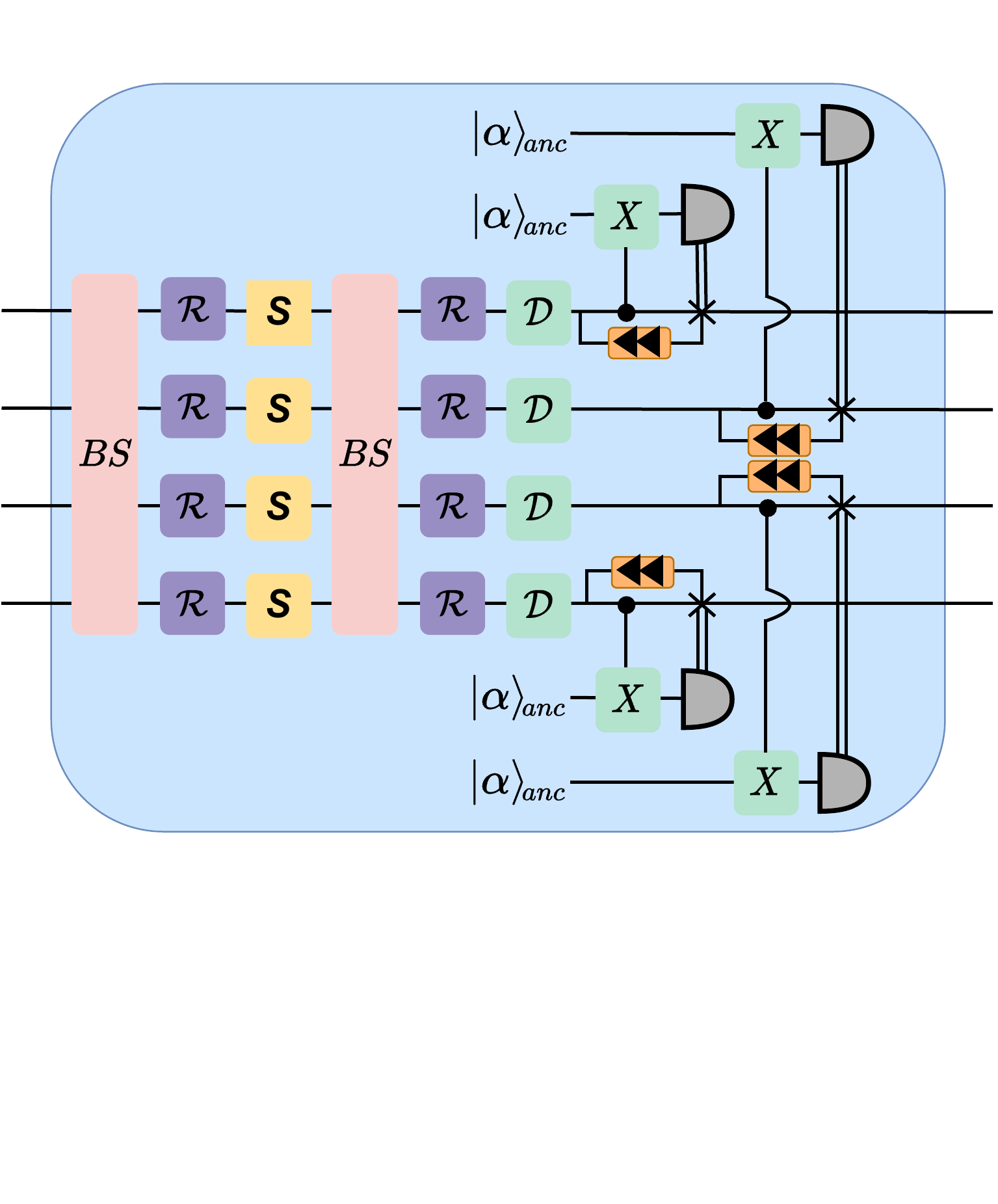}
\caption{Quantum layer}
\label{fig:mnist_ql}
\end{subfigure}
\caption{MNIST hybrid NN. (a) Encoding layer of MNIST hybrid neural network that embeds the output of the classical layers into the quantum circuit. (b) Detailed architecture of quantum layer. }
\end{figure}

The network architecture we used is illustrated in Fig.~\ref{fig:mnist_arch} with details of the encoding layer in Fig.~\ref{fig:mnist_enc} and the quantum layers in Fig.~\ref{fig:mnist_ql}. The network consists of fully connected feed-forward classical neural network layers that take the input data, feeding into an encoding quantum layer, followed by regular quantum layers that can be repeated as needed. Each quantum layer comprises several primary qumodes, each representing a class of the MNIST dataset, as well as ancillary qumodes that implement non-linearity.

During training, we calculated the probability or accuracy of classification by considering the overlap of the final state of the circuit, which comprises only the primary qumodes, with the one-hot encoded ground truth value of the training data sample. We counted all probabilities corresponding to non-zero values of the Fock number of the correct class towards the accuracy, indicating a ``click" or ``non-click" on a detector. The loss was then calculated using Eq.\ \eqref{eq:loss_func}. For training, we used the Adam optimizer with a batch size of 16 and a decaying learning rate beginning at 0.001 and decreasing by a factor of 0.9 every 5,000 steps. During validation or testing runs, we interpreted the probabilities corresponding to each primary qumode as logits. The predicted class was determined by selecting the logit with the highest value.

To assess the versatility of our model, we performed multiple experiments, exploring different configurations of classical and quantum layers in the hybrid model. We considered a range of ratios, starting from 1 classical layer and 5 quantum layers (1:5) up to 5 classical layers and 1 quantum layer (5:1). In total, we examined 8 different layer ratios, including ratios such as 2:2, 2:6, and 2:8. The classical layers comprised 128 nodes each, except for the final layer (or only layer in the case of a single classical layer model), which had the same number of nodes as required by the quantum encoding layer. The latter is determined by the formula $7p-2$,
where $p$ denotes the number of classes or primary qumodes.

\begin{figure}
    \begin{subfigure}{0.4 \textwidth}
\centering
\includegraphics[width=0.8 \textwidth]{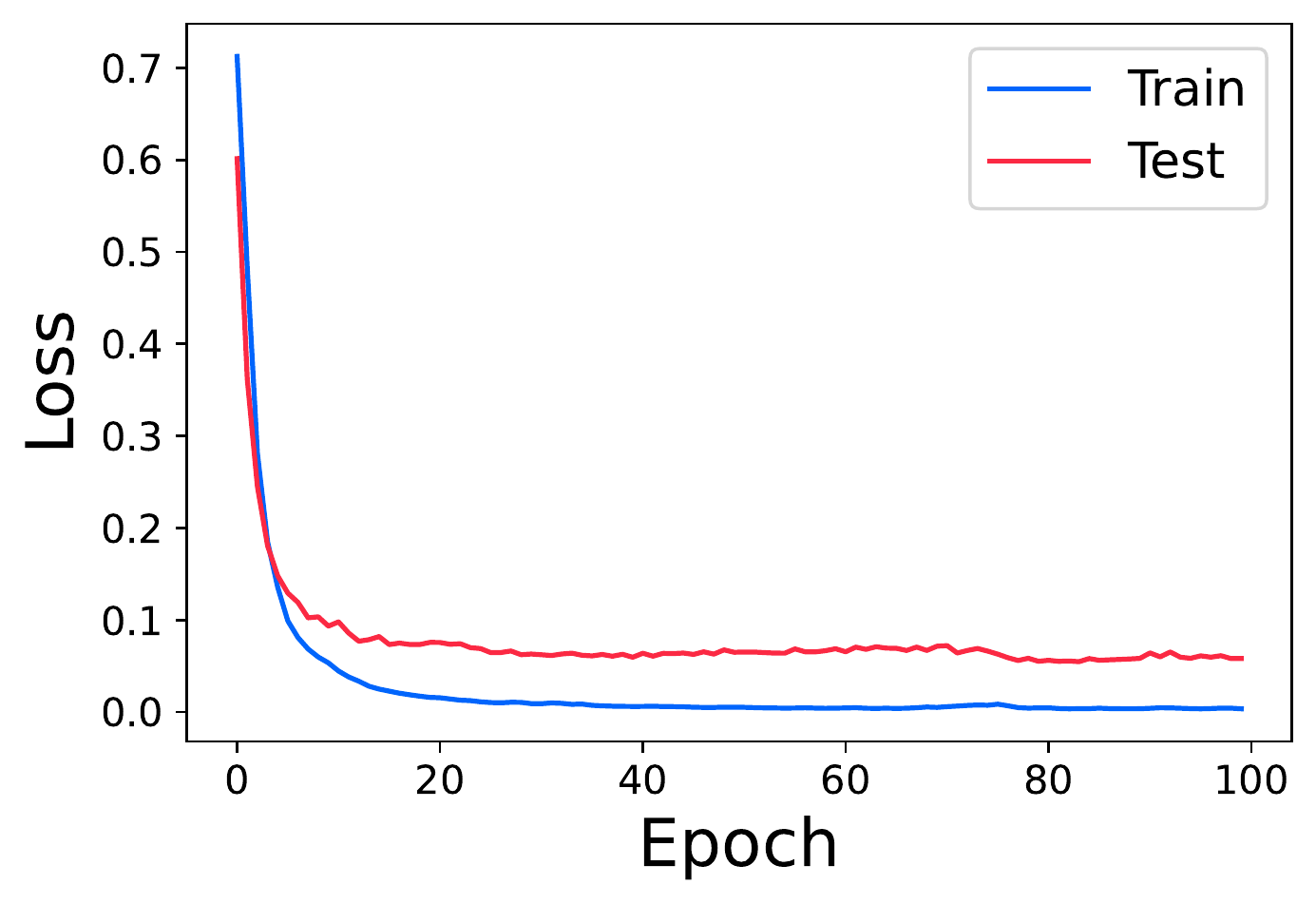}
\caption{1:5 Network}
\end{subfigure}
\centering
    \begin{subfigure}{0.4 \textwidth}
\includegraphics[width=0.8 \textwidth]{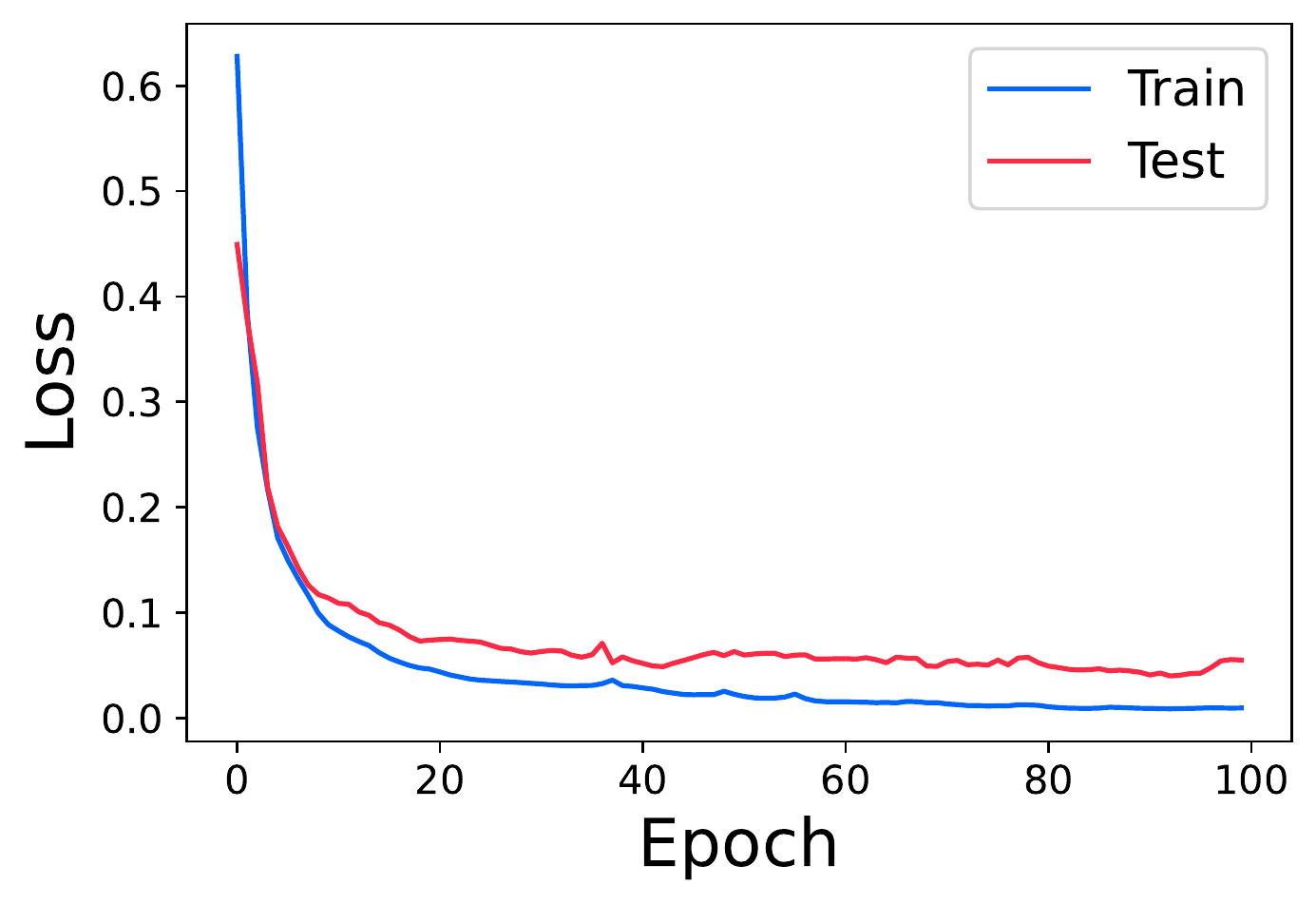}
\caption{3:3 Network}
\end{subfigure}
\centering
    \begin{subfigure}{0.4 \textwidth}
\includegraphics[width=0.8 \textwidth]{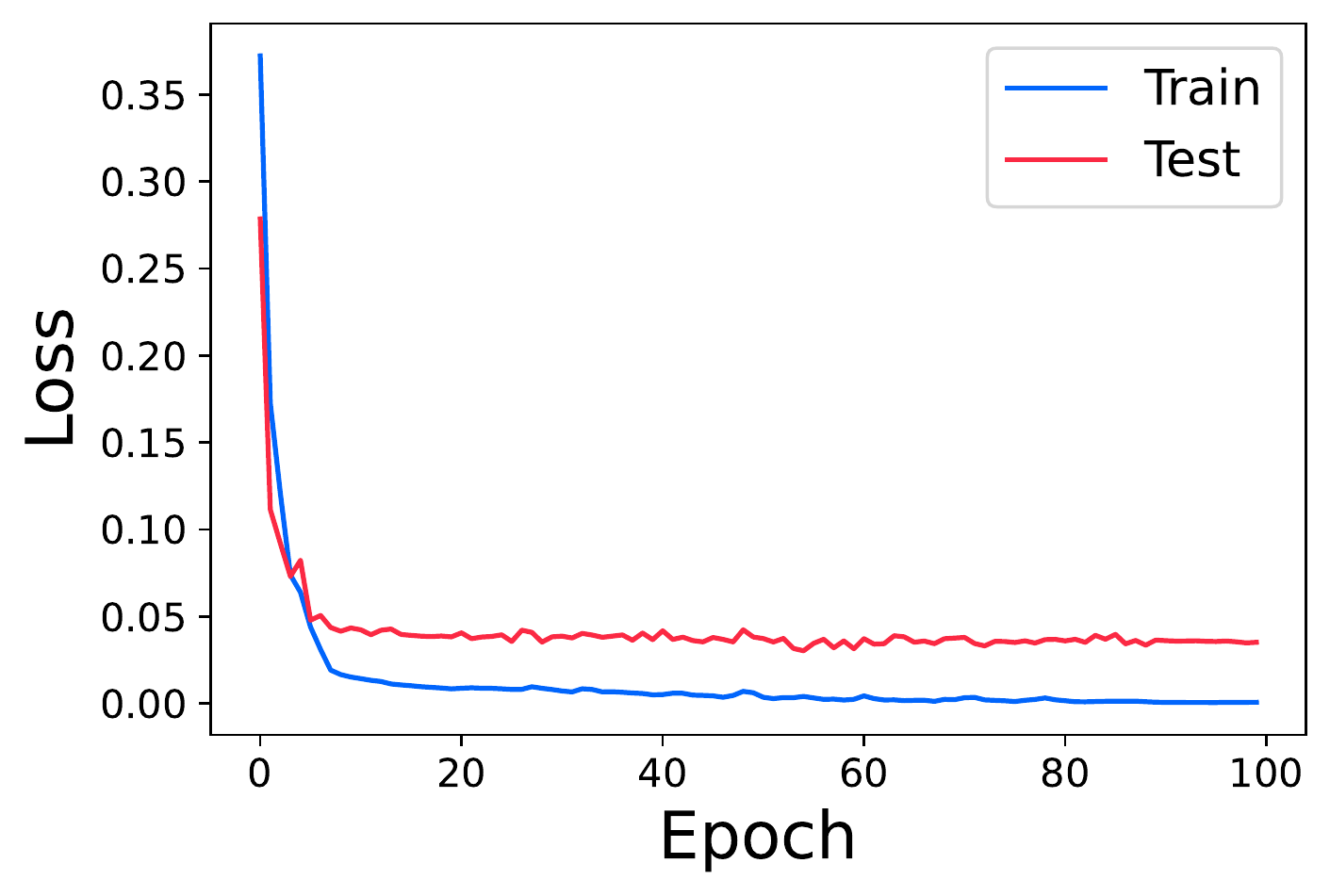}
\caption{5:1 Network}
\end{subfigure}
\caption{Training and testing loss plots for select ratios of classical and quantum layers, the first (second) number indicating the number of classical (quantum) layers. }\label{fig:mnist_ratio_loss}
\end{figure}

The loss function values over the epochs for select ratios of classical and quantum layers are displayed in Fig.~\ref{fig:mnist_ratio_loss}. All of our models converged within 100 epochs and demonstrated successful training and testing, achieving testing accuracies of $96.47\% \pm 0.86\%$. Interestingly, we did not observe a significant impact on changing the number of classical and quantum layers. Nevertheless, our hybrid networks demonstrated high levels of accuracy for the 4-class MNIST classification problem.

\section{Conclusion} \label{section:conclude}

We proposed CV QNN models that can be realized experimentally. We introduced a quantum circuit element that involves an ancillary qumode with a controlled-X, i.e. CX gate, on the primary qumode. For a good success rate, it relies on repeat-until-success measurements of the photon number of ancillary qumodes \cite{marshall2015repeat}. It offers a simple and feasible solution to introducing non-linearity using current photonic quantum hardware, considering the high complexity of implementing non-Gaussian operators experimentally. Our study demonstrated that our experimentally realizable circuit element could efficiently solve many machine-learning and quantum computation problems.

For instance, we created a CV quantum circuit that can prepare a single photon state with 99.9\% fidelity, a cat state with 99.8\% fidelity, and a GKP state of fidelity 93.9\%. In \cite{Arrazola_2019}, they performed the state preparation using CV QNNs and use the Kerr gate for non-linearity, making the entire process experimentally hard to achieve. However, our CV QNN can prepare the states in a way that can be realized experimentally. Although we could not achieve high fidelity for the GKP state because of the high computational requirements, we would need a higher cutoff dimension with more layers and optimization steps, leading to numerous training parameters for preparing such a complex state. However, it can be done, in principle, using high computational resources. GKP states could be a key factor in creating a scalable photonic fault-tolerant quantum computer \cite{Bourassa2021blueprintscalable}. 

We also developed models (from CV QNNs) capable of fitting functions using noisy data sets. We thoroughly analyzed the layers and noise effects on the curve fitting. As we increased the noise 5 times, the accuracy only went down by 24\%. Our model could still learn and accurately reproduce the curve's shape. This insight is valuable as our model can perform well even with noisy data sets. We encountered some challenges while attempting to approximate complex functions using our hybrid model for curve fitting. We discovered that these functions required a higher cutoff dimension and more layers to achieve a lower cost. This, in turn, necessitated additional computational resources for fitting more complex functions. Despite the issues, we successfully performed curve fitting on various functions like sine decay. It would be interesting to further study how quickly the QNN learns compared to its classical counterpart based on the complexity of the function and available data points for training. For the sine function we studied, the quantum and classical circuits reach the same accuracy after training, although there has been some work showing that with few training data, the quantum circuits could learn faster \cite{Caro22}. They do not provide an advantage over classical machine learning but it has been argued that quantum circuits can outperform their classical counterparts under certain assumptions. The importance of  data in quantum machine learning has also been studied; see, e.g., Ref.\ \cite{Huang21} where it was discussed how one could achieve quantum advantage based on chosen data and other machine learning techniques. 

The binary classification problem achieved high classification accuracy. The AUC score of 0.9 and accuracy of more than 95\% on a highly unbalanced data set. Similar work has been done to study fraud detection in credit card transactions using Quantum Support Vector Machine \cite{IBMFD}. They used a quantum-classical method to select the best features for the training process. They also focus on the importance of using quantum machine learning in selecting these features to improve the model's accuracy, which compliments the classical approach in finance. 

Also, image recognition done on the MNIST classification model can classify handwritten digits with up to 97\% accuracy. We did not observe a significant change in the results as we varied the mix of classical and quantum layers. Hence, future research could investigate the efficacy of quantum layers in hybrid neural networks and quantum neural networks in general. Quantum computation has the potential to provide exponential speedup over classical computation for certain problems, quantum layers can exploit this speedup to perform computations more efficiently than classical layers for tasks that can benefit from quantum algorithms. Quantum layers can also leverage the properties of superposition and entanglement to process and represent information in ways that are not possible with classical layers. Quantum layers can also be utilized as non-linear feature mappings that are challenging for classical layers. Quantum Feature Maps can be used to transform input data into higher dimensions for richer representation to later be utilized by classical neural networks, leading to potentially more accurate classification. Moreover, even though CV QGANs (quantum generative adversarial networks) have been previously studied (see, e.g., \cite{chang2021quantum}), it would be interesting to explore the performance of CV QGANs utilizing our proposed experimentally feasible setup. The proposed prototype of QGANs in \cite{chang2021quantum} requires the use of non-Gaussian gates within its quantum layers for both the quantum generator and the quantum discriminator. Since our prescribed quantum layer only requires Gaussian gates, we can simulate the effectiveness of an experimentally viable QGAN. 

Another possible future direction would be comparing continuous- and discrete-variable (DV) quantum computing. A similar study has been done in \cite{Abbas20}, where the authors compared the expressibility of classical and quantum neural networks by calculating effective dimensions for different cases. They showed that quantum neural networks have higher effective dimensions and train faster than their classical counterparts. They also used the Fisher information spectrum to demonstrate the resilience of quantum neural networks in terms of barren plateaus and the problem of vanishing gradients. It would be interesting to perform a similar study with CV QNNs. One such study has been done on barren plateaus in bosonic variational circuits \cite{zhang2023energy}. They used an energy-dependent circuit to prepare Gaussian and number states. It would be interesting to extend this to other problems in quantum machine learning by calculating effective dimensions for different models. CV QNNs have shown some advantage over their DV counterparts in terms of required resources, thus a study of the performance of CV vs.\ DV QNNs would be of interest. 

In conclusion, with our proposed CV quantum algorithm, we have obtained promising results in solving a wide range of machine-learning problems. The nonlinear quantum circuit element we introduced, which was based on an earlier proposal for universal CV quantum computing \cite{marshall2015repeat}, offers an experimentally feasible solution to introducing non-linearity using current photonic quantum hardware, avoiding the high complexity in experimentally realizing non-Gaussian operators.


\acknowledgements
Research funded by the National Science Foundation under award DGE-2152168. 
A portion of the computation for this work was performed on the University of Tennessee Infrastructure for Scientific Applications and Advanced Computing (ISAAC) computational resources.
KYA was supported by MITRE's Quantum Horizon Program.\footnote{\copyright 2023 The MITRE Corporation. ALL RIGHTS RESERVED. Approved for public release. Distribution unlimited PR$\textunderscore$22$-$04067$-$3.}






\end{document}